\documentclass[journal]{IEEEtran}
\usepackage{amsmath,amsfonts}
\usepackage{graphicx}
\usepackage{algorithm,algorithmic}
\usepackage{amsthm}
\usepackage{subfigure,url}
\usepackage[noadjust]{cite}
\usepackage[usenames,dvipsnames]{color}

\theoremstyle{plain}
\newtheorem{lem}[]{Lemma}

\DeclareMathOperator{\Tr}{Tr}
\newcommand{\Fopt}{\mathbf{F}_\mathrm{opt}}
\newcommand{\FRF}{\mathbf{F}_\mathrm{RF}}
\newcommand{\FBB}{\mathbf{F}_\mathrm{BB}}
\newcommand{\FDD}{\mathbf{F}_\mathrm{DD}}

\newcommand\relphantom[1]{\mathrel{\phantom{#1}}}
\ifCLASSINFOpdf
\else
\fi
\begin{document}
%
\title{Alternating Minimization Algorithms for Hybrid Precoding in Millimeter Wave MIMO Systems}
%
%
%

\author{Xianghao~Yu,~\IEEEmembership{Student Member,~IEEE}, 
	  Juei-Chin~Shen,~\IEEEmembership{Member,~IEEE}, 
        Jun~Zhang,~\IEEEmembership{Senior Member,~IEEE}, 
        and Khaled~B.~Letaief,~\IEEEmembership{Fellow,~IEEE}
\thanks{
Manuscript received June 1, 2015; revised November 6, 2015; accepted January 13, 2016. This work was supported by the Hong Kong Research Grants Council under Grant No. 610212. The guest editor coordinating the review of this paper and approving it for publication was Dr. Robert W. Heath.

This work was presented in part at IEEE Global Communications Conference, San Diego, CA, Dec. 2015 \cite{mine}.

X. Yu, J. Zhang, and K. B. Letaief are with the Department of Electronic and Computer Engineering, the Hong Kong University of Science and Technology (HKUST), Clear Water Bay, Kowloon, Hong Kong (email: \{xyuam, eejzhang, eekhaled\}@ust.hk).
K. B. Letaief is also with Hamad Bin Khalifa University, Qatar (email: kletaief@hbku.edu.qa).

J.-C. Shen is with MediaTek Inc., Hsinchu City 30078, Taiwan (e-mail: jc.shen@mediatek.com).}
}

\maketitle

\begin{abstract}
Millimeter wave (mmWave) communications has been regarded as a key enabling technology for 5G networks, as it offers orders of magnitude greater spectrum than current cellular bands.
In contrast to conventional multiple-input-multiple-output (MIMO) systems, precoding in mmWave MIMO cannot be performed entirely at baseband using digital precoders, as only a limited number of signal mixers and analog-to-digital converters (ADCs) can be supported considering their cost and power consumption.
As a cost-effective alternative, a hybrid precoding transceiver architecture, combining a digital precoder and an analog precoder, has recently received considerable attention. However, the optimal design of such hybrid precoders has not been fully understood. In this paper, treating the hybrid precoder design as a matrix factorization problem, effective alternating minimization (AltMin) algorithms will be proposed for two different hybrid precoding structures, i.e., the fully-connected and partially-connected structures. In particular, for the fully-connected structure, an AltMin algorithm based on manifold optimization is proposed to approach the performance of the fully digital precoder, which, however, has a high complexity. Thus, a low-complexity AltMin algorithm is then proposed, by enforcing an orthogonal constraint on the digital precoder. Furthermore, for the partially-connected structure, an AltMin algorithm is also developed with the help of semidefinite relaxation.
For practical implementation, the proposed AltMin algorithms are further extended to the broadband setting with orthogonal frequency division multiplexing (OFDM) modulation.
Simulation results will demonstrate significant performance gains of the proposed AltMin algorithms over existing hybrid precoding algorithms. Moreover, based on the proposed algorithms, simulation comparisons between the two hybrid precoding structures will provide valuable design insights.
\end{abstract}


%
\IEEEpeerreviewmaketitle

\section{Introduction}
%
%
%
%
\IEEEPARstart{T}{he} capacity of wireless networks has to exponentially increase to meet the explosive demands for high-data-rate multimedia access. In particular, the upcoming 5G networks aim at carrying out the projected 1000X increase in capacity by 2020 \cite{andrews2014will}. One way to boost the capacity is to improve the spectral efficiency through physical layer techniques, such as massive multiple-input-multiple-output (MIMO) and advanced channel coding \cite{hoydis2013massive}. Further improvement in area spectral efficiency can be achieved by network densification, such as deploying small cells \cite{li2013throughput,6812291} and allowing device-to-device (D2D) communications \cite{doppler2009device}, and enabling advanced cooperation, such as Cloud-RANs \cite{7143330,6786060}. Nevertheless, the spectrum crunch in current cellular systems brings a fundamental bottleneck for the further capacity increase. Thus, it is critical to exploit underutilized spectrum bands, including the bands that have not been used for cellular communications yet.

Millimeter wave (mmWave) bands from 30 GHz to 300 GHz, previously only considered for outdoor point-to-point
backhaul links \cite{hur2013millimeter} or for carrying indoor high-resolution multimedia streams \cite{torkildson2011indoor}, have now been put forward as a prime candidate for new spectrum in 5G cellular systems, with the potential bandwidth reaching 10 GHz. This view is supported by recent experiments in New York City that demonstrated the feasibility of mmWave outdoor cellular communications \cite{akdeniz2013millimeter,rappaport2014millimeter}. Originally, the main obstacles for the success of mmWave cellular systems are the huge path loss and rain attenuation, as a result of the ten-fold increase of the carrier frequency \cite{akdeniz2013millimeter}.
Thanks to the small wavelength of mmWave signals, mmWave MIMO precoding can leverage large-scale antennas at transceivers to provide significant beamforming gains to combat the path loss and to synthesize highly directional beams. Moreover, spectral efficiency can be further increased by transmitting multiple data streams via spatial multiplexing.

For traditional MIMO systems, precoding is typically accomplished at baseband through digital precoders, which can adjust both the magnitude and phase of the signals. However, fully digital precoding demands radio frequency (RF) chains, including signal mixers and analog-to-digital converters (ADCs), comparable in number to the antenna elements. While the small wavelengths of mmWave frequencies facilitate the use of a large number of antenna elements, the prohibitive cost and power consumption of RF chains make digital precoding infeasible. Given such unique constraints in mmWave MIMO systems, a hybrid precoding architecture has recently received much consideration, which only requires a small number of RF chains interfacing between a low-dimensional digital precoder and a high-dimensional analog precoder \cite{el2014spatially}. As the analog precoders are still of high dimension, it is impractical to implement them in the RF domain with power-hungry variable voltage amplifiers (VGAs) \cite{rappaport2014millimeter}. This heuristic leads to a rule of thumb, i.e., realizing analog precoders with low-cost phase shifters at the expense of sacrificing the ability to change the magnitude of the RF signals. 

According to the mapping from RF chains to antennas, which determines the number of phase shifters in use, the hybrid precoding transceiver architectures can be categorized into the fully-connected and partially-connected structures, as illustrated in Fig. \ref{fullconnected} and Fig. \ref{subarray}, respectively. The former structure enjoys the full beamforming gain for each RF chain with a natural combination between RF chains and antenna elements, i.e., each RF chain is connected to all antennas. On the other hand, sacrificing some beamforming gain, the partially-connected structure significantly reduces the hardware implementation complexity by connecting each RF chain only with part of the antennas.

In \cite{el2014spatially}, it has been pointed out that maximizing the spectral efficiency of mmWave systems can be approximated by minimizing the Euclidean distance between hybrid precoders and the fully digital precoder. This renders the hybrid precoder design as a matrix factorization problem with unit modulus constraints imposed by the phase shifters. Although significant amounts of research efforts have been invested in solving various matrix factorization problems in recent years \cite{jain2013low,drineas2005nystrom}, with the unique unit modulus constraints, the optimal design of hybrid precoders remains unknown. Existing works often add some extra constraints on analog precoders to simplify the analog part design with unit modulus constraints, which will cause performance loss. This motivates us to reconsider the hybrid precoder design or, in other words, the matrix factorization problem with unit modulus constraints on analog precoders. In particular, a better way to deal with the unit modulus constraint deserves further delicate investigations.

In this paper, by adopting alternating minimization (AltMin) as the main design approach, we will propose different hybrid precoding algorithms to approach the performance of the optimal fully digital precoder. Based on the principle of alternating minimization, three novel algorithms will be proposed to find effective hybrid precoding solutions for the fully-connected and partially-connected structures.

\subsection{Related Works}
Hybrid precoding is a newly-emerged technique in mmWave MIMO systems 
\cite{roh2014millimeter,sun2014mimo,alkhateeb2014channel,wang2015multi,rangan2014millimeter}.
So far the main efforts are on the fully-connected structure \cite{el2014spatially,lee2014hybrid,6884253,kim2013mse,brady2013beamspace,7248509,zhang2014achieving,liang2014low,wang2014joint}.
Orthogonal matching pursuit (OMP) is the most widely used algorithm, which often offers reasonably good performance. This algorithm requires the columns of analog precoding matrix to be picked from certain candidate vectors, such as array response vectors of the channel \cite{el2014spatially,lee2014hybrid,6884253}, and discrete Fourier transform (DFT) beamformers \cite{kim2013mse,brady2013beamspace}. Hence, the OMP-based hybrid precoder design can be viewed as a sparsity constrained matrix reconstruction problem. Though the design problem is greatly simplified in this way, restricting the space of feasible analog precoding solutions inevitably causes some performance loss. Additionally, extra overhead will be brought up for acquiring the information of array response vectors in advance. More recent attention has mainly focused on reducing the computation complexity of the OMP algorithm \cite{lee2014hybrid,7248509}, e.g., by reusing the matrix inversion result in each iteration.

There are works investigating some special hybrid precoding systems. In \cite{zhang2014achieving}, an optimal hybrid precoder design in a special case was identified, i.e., when the number of RF chains is at least twice that of the data streams. However, the optimal solution for the general case is unknown.
The authors of \cite{wang2014joint} investigated VGA-enabled hybrid precoding according to different design criteria. By removing VGAs from the RF domain, low-power analog precoders with phase shifters were also considered in \cite{wang2014joint}, whose phases are heuristically extracted from those of the VGA-enabled solution.

On the other hand, much less attention has been paid on the partially-connected structure \cite{singh2014feasibility,6824962,7248508,han2015large,zhang2015massive,el2013multimode}. 
In \cite{singh2014feasibility,6824962}, codebook-based design of hybrid precoders was presented for narrowband and orthogonal frequency division multiplexing (OFDM) systems, respectively.
Although the codebook-based design enjoys a low complexity, there will be certain performance loss, and it is not clear how much performance gain can be further obtained. 
By utilizing the idea of successive interference cancellation (SIC), an iterative hybrid precoding algorithm for the partially-connected structure was proposed in \cite{7248508}. The algorithm is established based on the assumption that the digital precoding matrix is diagonal, which means that the digital precoder only allocates power to different data streams, and the number of RF chains should be equal to that of the data streams.
 However, using only analog precoders to provide beamforming gains is obviously a suboptimal strategy \cite{7248508,han2015large}, which also deviates from the motivation of hybrid precoding.
So far there is no study directly optimizing the hybrid precoders without extra constraints in the partially-connected structure, which will be pursued in this paper.
\subsection{Contributions}
In this paper, we investigate the hybrid precoder design in mmWave MIMO systems. We will adopt alternating minimization (AltMin) as the main design principle, which helps decouple the precoder design problem into two subproblems, i.e., the analog and digital precoder design. The proposed AltMin algorithms will alternately optimize the digital precoder and the analog precoder. Our major contributions are summarized as follows:
\begin{itemize}
\item For the fully-connected structure, we shall show that the unit modulus constraints of the analog precoder define a Riemannian manifold. We will thus propose a manifold optimization based AltMin (MO-AltMin) algorithm.
This algorithm does not need any pre-determined candidate set for the analog precoder, and it is the first attempt to directly solve the hybrid precoder design problem under the unit modulus constraints.
\item By imposing an orthogonal property of the digital precoder, we then develop an AltMin algorithm using phase extraction (PE-AltMin) as a low-complexity counterpart of the MO-AltMin algorithm, which will also be more practical for implementation. 
\begin{figure*}[htbp]
\centering
\subfigure[MmWave MIMO transmitter architecture.]{
\includegraphics[height=6cm]{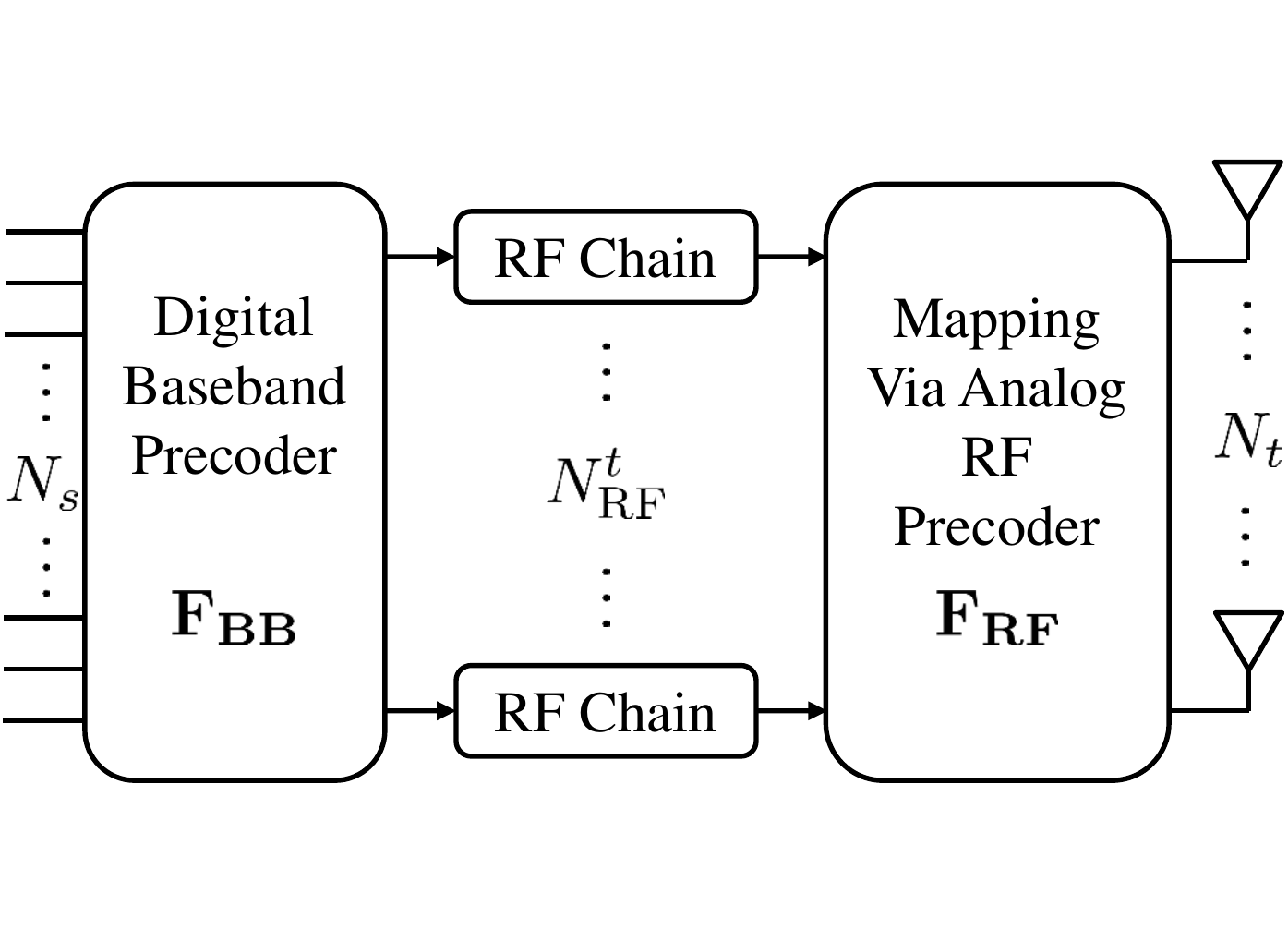}\label{systemmodel}
}
\subfigure[The mapping strategy for the fully-connected structure.]{
\includegraphics[height=6cm]{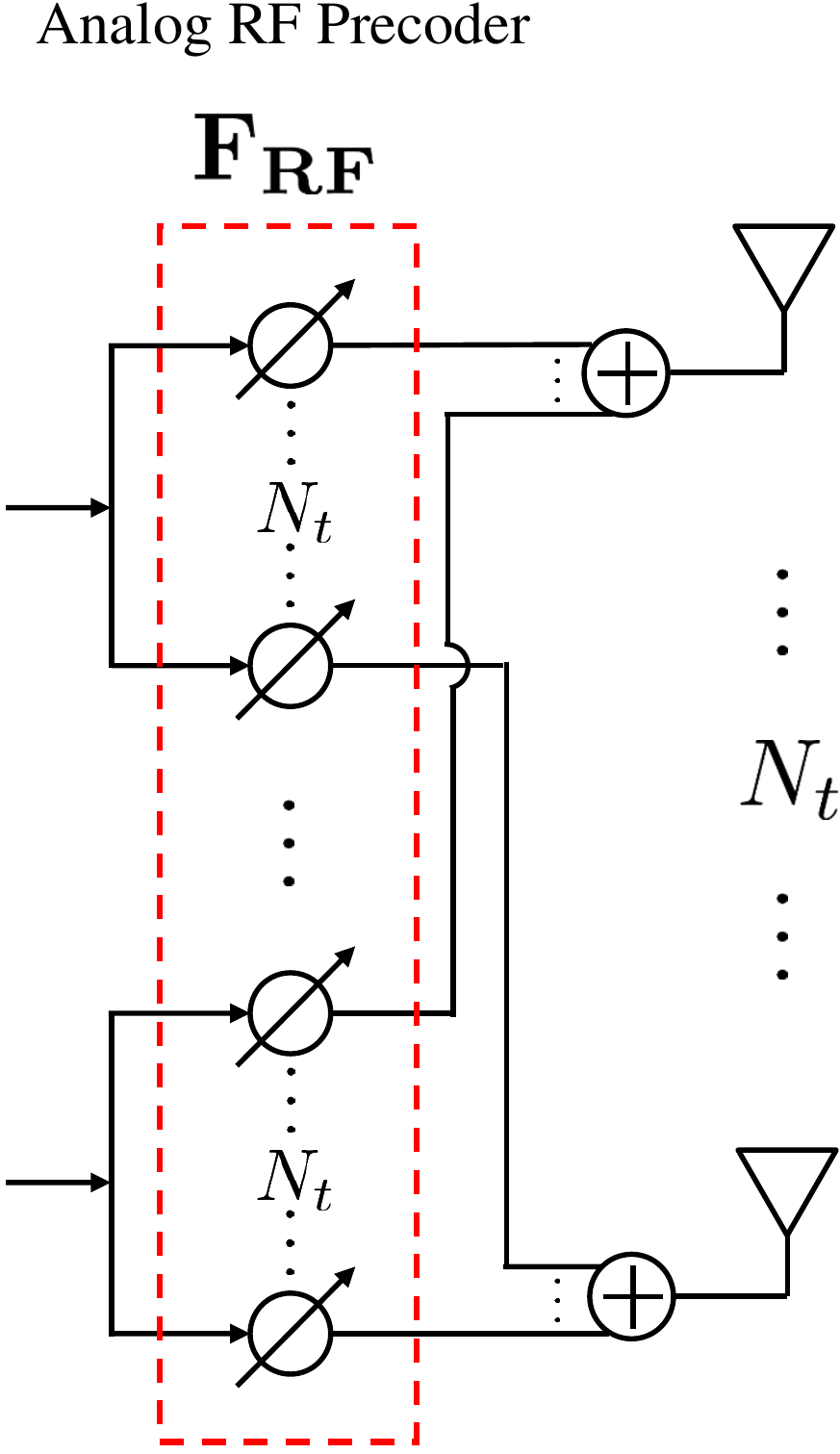}\label{fullconnected}
}
\subfigure[The mapping strategy for the partially-connected structure.]{
\includegraphics[height=6cm]{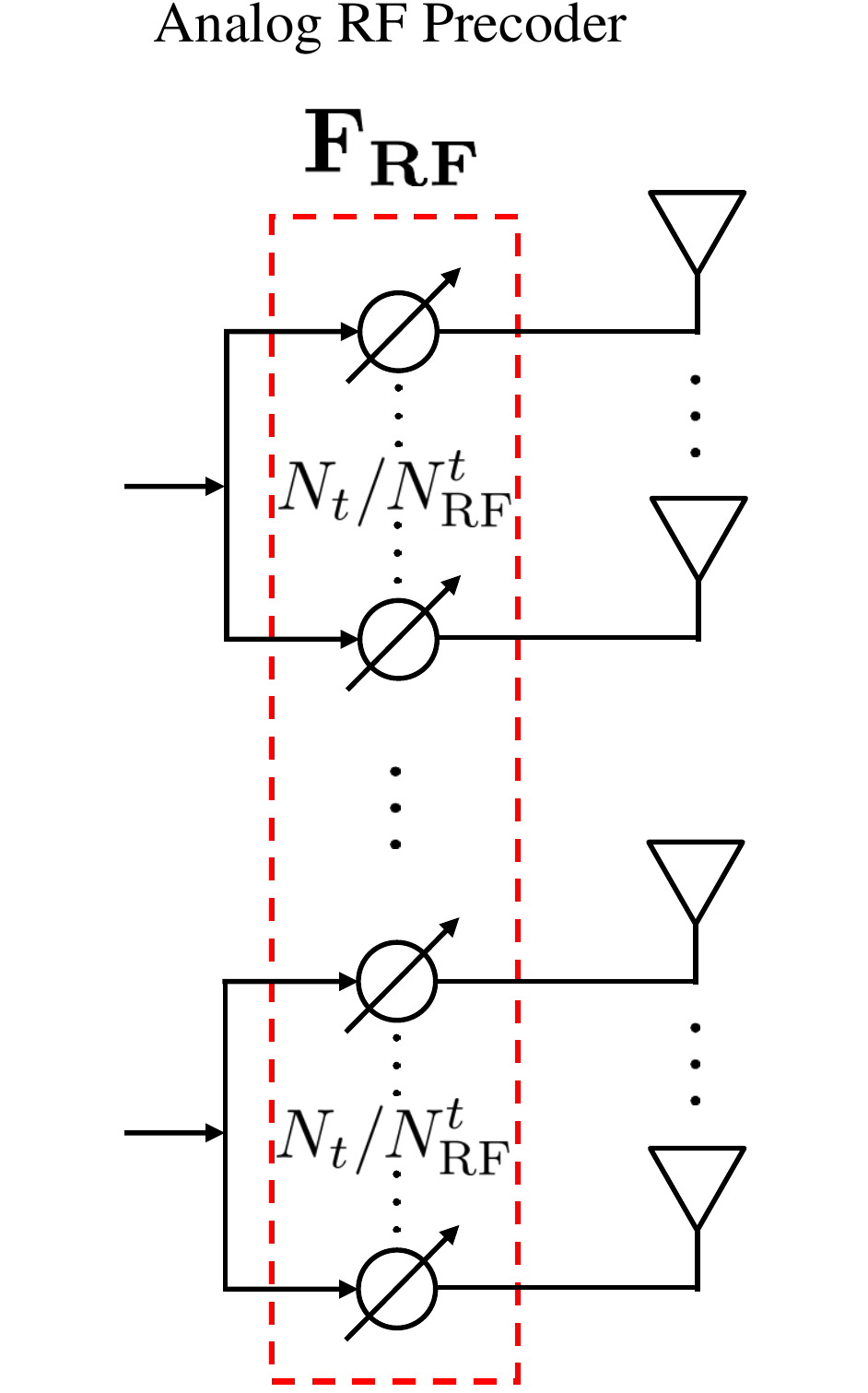}\label{subarray}
}
\caption{Two structures of hybrid precoding in mmWave MIMO systems using different mapping strategies: each RF chain is connected to $N_t$ antennas in (b) and to $N_t/N_\mathrm{RF}^t$ antennas in (c).}
\end{figure*}
\item For the partially-connected structure, we propose a semidefinite relaxation based AltMin (SDR-AltMin) algorithm. This algorithm effectively designs the hybrid precoders by offering optimal solutions for both subproblems of analog and digital precoders in each alternating iteration, and it is the first effort directly optimizing the hybrid precoders in such a structure.
\item The three proposed AltMin Algorithms can be generally applied to both narrowband and broadband OFDM systems. Simulation results will demonstrate that the MO-AltMin algorithm efficiently identifies a near-optimal solution, while the PE-AltMin algorithm with practical computational complexity outperforms the existing OMP algorithm.
\item With the proposed AltMin algorithms, extensive comparisons are provided to reveal valuable design insights.
In particular, the proposed AltMin algorithms for the fully-connected structure can help to approach the performance of the fully digital precoder as long as the number of RF chains is comparable to the number of data streams, which cannot be achieved by the widely applied OMP algorithm.
On the other hand, the SDR-AltMin algorithm for the partially-connected structure provides significant gains over analog beamforming. 
Furthermore, by taking advantage of its low-complexity hardware implementation, the partially-connected structure provides a higher energy efficiency than the fully-connected one with a relatively large number of RF chains implemented at transceivers.
\end{itemize}

Thus, our results firmly establish the effectiveness of the alternating minimization as a key design methodology for hybrid precoder design in mmWave MIMO systems.
\subsection{Organization}
The remainder of this paper is organized as follows. We shall introduce the system model and channel model, followed by the problem formulation in Section II. Two alternating minimization algorithms for the fully-connected structure are demonstrated in Section III and Section IV, respectively. In Section V, the hybrid precoder design for the partially-connected structure is investigated. Extensions of the proposed algorithms to OFDM systems are provided in Section VI, and simulation results will be presented in Section VII. Finally, we will conclude this paper in Section VIII.

\subsection{Notations}
The following notations are used throughout this paper. $\mathbf{a}$ and $\mathbf{A}$ stand for a column vector and a matrix, respectively; $\mathbf{A}_{i,j}$ is the entry on the $i$th row and $j$th column of $\mathbf{A}$; The conjugate, transpose and conjugate transpose of $\mathbf{A}$ are represented by $\mathbf{A}^*$, $\mathbf{A}^T$ and $\mathbf{A}^H$; $\det(\mathbf{A})$ and $\left\|\mathbf{A}\right\|_F$ denote the determinant and Frobenius norm of $\mathbf{A}$; $\mathbf{A}^{-1}$ and $\mathbf{A}^{\dag}$ are the inverse and Moore-Penrose pseudo inverse of $\mathbf{A}$; $\mathrm{Tr}(\mathbf{A})$ and $\mathrm{vec}(\mathbf{A})$ indicate the trace and vectorization; Expectation and the real part of a complex variable is noted by $\mathbb{E}[\cdot]$ and $\Re[\cdot]$; $\circ$ and $\otimes$ denote the Hadamard and Kronecker products between two matrices.

\section{System Model and Problem Formulation}
In this section, we will first present the system model and channel model of the considered mmWave MIMO system, and then formulate the hybrid precoding problem.
\subsection{System Model}\label{IIA}
Consider a single-user mmWave MIMO system\footnote{The receiver side is omitted due to space limitation. More details can be found in \cite{el2014spatially}.} as shown in Fig. \ref{systemmodel}, where $N_s$ data streams are sent and collected by $N_t$ transmit antennas and $N_r$ receive antennas.
The numbers of RF chains at the transmitter and receiver are respectively denoted as $N_\mathrm{RF}^t$ and $N_\mathrm{RF}^r$, which are subject to constraints $N_s\le N_\mathrm{RF}^t\le N_t$ and $N_s\le N_\mathrm{RF}^r\le N_r$.

The transmitted signal can be written as $\mathbf{x}=\mathbf{F}_\mathrm{RF}\mathbf{F}_\mathrm{BB}\mathbf{s}$, where $\mathbf{s}$ is the $N_s\times 1$ symbol vector such that $\mathbb{E}\left[\mathbf{ss}^H\right]=\frac{1}{N_s}\mathbf{I}_{N_s}$. The hybrid precoders consist of an $N_\mathrm{RF}^t\times N_s$ digital baseband precoder $\mathbf{F}_\mathrm{BB}$ and an $N_t\times N_\mathrm{RF}^t$ analog RF precoder $\mathbf{F}_\mathrm{RF}$. The normalized transmit power constraint is given by $\left\|\mathbf{F}_\mathrm{RF}\mathbf{F}_\mathrm{BB}\right\|_F^2=N_s$. For simplicity, we first consider a narrowband block-fading propagation channel, while the extension to broadband OFDM systems will be treated in Section VI. Thus, 
the received signal after decoding processing is given as
\begin{equation}
\mathbf{y}=\sqrt{\rho}\mathbf{W}_\mathrm{BB}^H\mathbf{W}_\mathrm{RF}^H\mathbf{H}\mathbf{F}_\mathrm{RF}\mathbf{F}_\mathrm{BB}\mathbf{s}+\mathbf{W}_\mathrm{BB}^H\mathbf{W}_\mathrm{RF}^H\mathbf{n},
\end{equation}
where $\rho$ stands for the average received power, $\mathbf{H}$ is the channel matrix, $\mathbf{W}_\mathrm{BB}$ is the $N_\mathrm{RF}^r\times N_s$ digital baseband decoder, $\mathbf{W}_\mathrm{RF}$ is the $N_r\times N_\mathrm{RF}^r$ analog RF decoder at the receiver, and $\mathbf{n}$ is the noise vector of independent and identically distributed (i.i.d.) $\mathcal{CN}(0,\sigma_n^2)$ elements.
In this paper, we assume that perfect channel state information (CSI) is known at both the transmitter and receiver. In practice, CSI can be accurately and efficiently obtained by channel estimation \cite{alkhateeb2014channel} at the receiver and further shared at the transmitter with effective feedback techniques \cite{el2014spatially,1468321}.
The achievable spectral efficiency when transmitted symbols follow a Gaussian distribution can be expressed as
\begin{equation}\label{spectralefficiency}
\begin{split}
R=\log\det\bigg(\mathbf{I}_{N_s}+\frac{\rho}{\sigma_n^2 N_s}{{{\left( {{{\bf{W}}_{{\rm{RF}}}}{{\bf{W}}_{{\rm{BB}}}}} \right)}^\dag }{\bf{H}}{{\bf{F}}_{{\rm{RF}}}}{{\bf{F}}_{{\rm{BB}}}}}\\
\times{{\bf{F}}_{{\rm{BB}}}^H{\bf{F}}_{{\rm{RF}}}^H{{\bf{H}}^H}\left( {{{\bf{W}}_{{\rm{RF}}}}{{\bf{W}}_{{\rm{BB}}}}} \right)}\bigg).
\end{split}
\end{equation}
Furthermore, the analog precoders are implemented with phase shifters, which can only adjust the phases of the signals. Thus, all the nonzero entries of $\mathbf{F}_\mathrm{RF}$ and $\mathbf{W}_\mathrm{RF}$ should satisfy the unit modulus constraints, namely $|({\mathbf{F}_\mathrm{RF}})_{i,j}|=|({\mathbf{W}_\mathrm{RF}})_{i,j}|=1$ for nonzero elements.

According to different signal mapping strategies from RF chains to antennas, the transceiver architecture can be categorized into the fully-connected and partially-connected hybrid precoding structures, as illustrated in Fig. \ref{fullconnected} and Fig. \ref{subarray}. For the fully-connected structure, the output signal of each RF chain is sent to all the antennas through phase shifters, while the partially-connected structure only has $N_t/N_\mathrm{RF}^t$ antennas connected to each RF chain. Thus, the fully-connected structure enjoys full beamforming gain for each RF chain with a natural combination between RF chains and antennas, whereas the hardware implementation complexity is lower in the partially-connected one by sacrificing some beamforming gain for each RF chain. The structures of $\FRF$s and $\mathbf{W}_\mathrm{RF}$s will vary for different structures, which will be discussed in the following sections in detail. 

\subsection{Channel Model}
Due to high free-space path loss, the mmWave propagation environment is well characterized by a clustered channel model, i.e., the Saleh-Valenzuela model \cite{rappaport2014millimeter}. This model depicts the mmWave channel matrix as
\begin{equation}\label{channelmodel}
\mathbf{H}=\sqrt{\frac{N_tN_r}{N_{cl}N_{ray}}}\sum_{i=1}^{N_{cl}}\sum_{l=1}^{N_{ray}}{\alpha_{il}\mathbf{a}_r(\phi_{il}^r,\theta_{il}^r)\mathbf{a}_t(\phi_{il}^t,\theta_{il}^t)^H},
\end{equation}
where $N_{cl}$ and $N_{ray}$ represent the number of clusters and the number of rays in each cluster, and $\alpha_{il}$ denotes the gain of the $l$th ray in the $i$th propagation cluster. We assume that $\alpha_{il}$ are i.i.d. and follow the distribution $\mathcal{CN}(0,\sigma_{\alpha,i}^2)$ and $\sum_{i=1}^{N_{cl}}{\sigma_{\alpha,i}^2=\hat\gamma}$, which is the normalization factor to satisfy $\mathbb{E}\left[\left\|\mathbf{H}\right\|_F^2\right]=N_tN_r$. In addition, $\mathbf{a}_r(\phi_{il}^r,\theta_{il}^r)$ and $\mathbf{a}_t(\phi_{il}^t,\theta_{il}^t)$ represent the receive and transmit array response vectors, where $\phi_{il}^r$($\phi_{il}^t$) and $\theta_{il}^r$($\theta_{il}^t$) stand for azimuth and elevation angles of arrival and departure, respectively. In this paper, we consider the uniform square planar array (USPA) with $\sqrt{N}\times\sqrt{N}$ antenna elements. Therefore, the array response vector corresponding to the $l$th ray in the $i$th cluster can be written as
\begin{equation}
\begin{split}
\mathbf{a}(\phi_{il},\theta_{il})=\frac{1}{\sqrt{N}}\bigg[1,\cdots,e^{{j\frac{2\pi}{\lambda}d(p \sin\phi_{il}\sin\theta_{il}+q\cos\theta_{il})}},\\
\cdots,e^{{j\frac{2\pi}{\lambda}d\left(\left(\sqrt{N}-1\right) \sin\phi_{il}\sin\theta_{il}+\left(\sqrt{N}-1\right)\cos\theta_{il}\right)}}\bigg]^T,
\end{split}
\end{equation}
where $d$ and $\lambda$ are the antenna spacing and the signal wavelength, and $0\le p< \sqrt{N}$ and $0\le q< \sqrt{N}$ are the antenna indices in the 2D plane. While this channel model will be used in simulations, our precoder design is applicable to more general models.

\subsection{Problem Formulation}
As shown in \cite{el2014spatially,lee2014hybrid}, the design of precoders and decoders can be separated into two subproblems, i.e., the precoding and decoding problems. They have similar mathematical formulations except that there is an extra power constraint in the former. Therefore, we will mainly focus on the precoder design in the remaining part of this paper and the algorithms proposed in this paper can be equally applied for the decoder. The corresponding problem formulation is given by
\begin{equation}\label{problemformulation}
\begin{aligned}
&\underset{\mathbf{F}_\mathrm{RF},\mathbf{F}_\mathrm{BB}}{\mathrm{minimize}} && \left\Vert \mathbf{F}_\mathrm{opt}-\mathbf{F}_\mathrm{RF}\mathbf{F}_\mathrm{BB}\right\Vert _F\\
&\mathrm{subject\thinspace to}&&
\begin{cases}
\FRF\in\mathcal{A}\\
\left\|\mathbf{F}_\mathrm{RF}\mathbf{F}_\mathrm{BB}\right\|_F^2=N_s,
\end{cases}
\end{aligned}
\end{equation}
where $\mathbf{F}_\mathrm{opt}$ stands for the optimal fully digital precoder, while $\mathbf{F}_\mathrm{RF}$ and $\mathbf{F}_\mathrm{BB}$ are the analog and digital precoders to be optimized. Additionally, $\mathcal{A}\in\{\mathcal{A}_f,\mathcal{A}_p\}$ is the feasible set of the analog precoder induced by the unit modulus constraints, which will be distinct for different hybrid precoding structures.

It has been shown in \cite{el2014spatially} that minimizing the objective function in \eqref{problemformulation} approximately leads to the maximization of the spectral efficiency. It is also intuitively true that the optimal hybrid precoders should be sufficiently ``close'' to the unconstrained optimal digital precoder. In addition, the unconstrained optimal precoder and decoder are comprised of the first $N_s$ columns of $\mathbf{V}$ and $\mathbf{U}$ respectively, which are unitary matrices derived from the channel's singular value decomposition (SVD), i.e., $\mathbf{H}=\mathbf{U\Sigma V}^H$.

We will mainly treat problem \eqref{problemformulation} as a matrix factorization problem, for which alternating minimization will be adopted as the main tool. Alternating minimization represents a widely applicable and empirically successful approach for the optimization problems involving different subsets of variables. It has been successfully applied to many applications such as matrix completion \cite{jain2013low}, phase retrieval \cite{NIPS2013_5041}, image reconstruction \cite{wang2008new}, blind deconvolution \cite{chan2000convergence} and non-negative matrix factorization \cite{kim2008nonnegative}. In this paper, the problem formulation \eqref{problemformulation} is intrinsically a matrix factorization problem involving two matrix variables $\FRF$ and $\FBB$. However, jointly optimizing these two variables is highly complicated due to the element-wise unit modulus constraints of $\FRF$. By decoupling the optimization of these two variables, alternating minimization stands out as an efficient method to obtain an effective solution. With the principle of alternating minimization, we will alternately solve for $\FRF$ and $\FBB$ while fixing the other, which will be the essential idea throughout this paper.

\section{Manifold Optimization Based Hybrid Precoding for the Fully-connected Structure}\label{III}
The fully-connected structure, in which each RF chain is connected to all the antenna elements, is frequently used in mmWave MIMO systems, as shown in Fig. \ref{fullconnected}. This structure restricts every entry in the analog precoding matrix to be unit modulus, and this element-wise constraint makes the precoder design problem intractable. In this section, by observing that the unit modulus constraints define a Riemannian manifold, we will propose an AltMin algorithm based on manifold optimization to directly solve \eqref{problemformulation}.

For the fully-connected structure, inspired by \cite{variable05}, the authors of \cite{zhang2014achieving} have shown that the Frobenius norm in \eqref{problemformulation} can be made exactly zero under the condition that $N_\mathrm{RF}^t\ge2N_s$. This means that the hybrid precoders can achieve the performance of the fully digital precoder in this special case, and the optimal hybrid precoders were obtained in \cite{zhang2014achieving}. Thus, we will focus on the region where $N_s\le N_\mathrm{RF}^t<2N_s$ in this paper.

\subsection{Digital Baseband Precoder Design}\label{S31}
We first consider to design the digital precoder $\mathbf{F}_\mathrm{BB}$ with a fixed analog precoder $\mathbf{F}_\mathrm{RF}$. Thus, problem \eqref{problemformulation} can be restated as
\begin{equation}\label{digital}
\begin{aligned}
&\underset{\mathbf{F}_\mathrm{BB}}{\mathrm{minimize}} && \left\Vert \mathbf{F}_\mathrm{opt}-\mathbf{F}_\mathrm{RF}\mathbf{F}_\mathrm{BB}\right\Vert _F,
\end{aligned}
\end{equation}
which has a well-known least squares solution given by
\begin{equation}\label{ls}
\mathbf{F}_\mathrm{BB}=\mathbf{F}_\mathrm{RF}^\dag\mathbf{F}_\mathrm{opt}.
\end{equation}

Note that the power constraint in \eqref{problemformulation} is temporarily removed, and it will be dealt with in Section \ref{C}. Nevertheless, the solution in \eqref{ls} has already offered a globally optimal solution to the counterpart design problem at the receiver side.

\subsection{Analog RF Precoder Design via Manifold Optimization}
For the fully-connected structure, the feasible set $\mathcal{A}_f$ of the analog precoder can be specified by $|{(\mathbf{F}_\mathrm{RF}})_{i,j}|=1$, as each RF chain is connected to all the antennas. In the next alternating step, we fix $\mathbf{F}_\mathrm{BB}$ and seek an analog precoder which optimizes the following problem\footnote{The square of the Frobenius norm makes the objective function quadratic and smooth, and will not affect the solution.}:
\begin{equation}\label{analog}
\begin{aligned}
&\underset{\mathbf{F}_\mathrm{RF}}{\mathrm{minimize}} && \left\Vert \mathbf{F}_\mathrm{opt}-\mathbf{F}_\mathrm{RF}\mathbf{F}_\mathrm{BB}\right\Vert _F^2\\
&\mathrm{subject\thinspace to} &&|({\mathbf{F}_\mathrm{RF}})_{i,j}|=1, \forall i,j.
\end{aligned}
\end{equation}
The main obstacles are the unit modulus constraints, which are intrinsically non-convex. To the best of the authors' knowledge, there is no general approach to solve \eqref{analog} optimally.
In the following, we will propose an effective manifold optimization algorithm to find a near-optimal solution of problem \eqref{analog}.
\begin{figure}[htbp]
\centering\includegraphics[width=6cm]{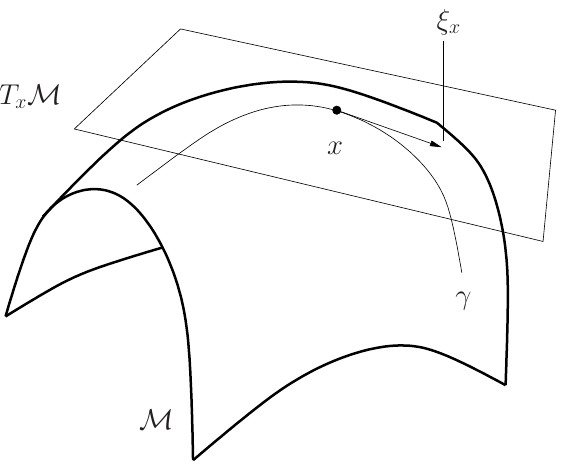}
\caption{The tangent space and tangent vector of a Riemannian manifold \cite{absil2009optimization}.}\label{m}
\end{figure}

We will start with some definitions and terminologies in manifold optimization. More background on manifolds and manifold optimization can be found in \cite{absil2009optimization,ma2012manifold,lee2012introduction}, and there are some recent applications in wireless communications \cite{yuanming}. As shown in Fig. \ref{m}, a \emph{manifold} $\mathcal{M}$ is a topological space that resembles a Euclidean space near each point \cite{lee2012introduction}. In other words, each point on a manifold has a neighborhood that is homeomorphic to the Euclidean space. The \emph{tangent space} $T_x\mathcal{M}$ at a given point $x$ on the manifold $\mathcal{M}$ is composed of the \emph{tangent vectors} $\xi_x$ of the curves $\gamma$ through the point $x$. In most applications, manifolds fall into a special category of topological manifold, namely, a \emph{Riemannian manifold}. A Riemannian manifold is equipped with an inner product defined on the tangent spaces $T_x\mathcal{M}$, called the \emph{Riemannian metric}, which allows one to measure distances and angles on manifolds. In particular, it is possible to use calculus on a Riemannian manifold with the Riemannian metric.

The rich geometry of Riemannian manifolds makes it possible to define gradients of cost functions. More importantly, optimization over a Riemannian manifold is locally analogous to that over a Euclidean space with smooth constraints. Therefore, a well-developed conjugate gradient algorithm in Euclidean spaces can find its counterpart on the specified Riemannian manifolds. In the following, we will briefly introduce this counterpart.

We first endow the complex plane $\mathbb{C}$ with the Euclidean metric
\begin{equation}
\left\langle {x_1,x_2} \right\rangle=\Re\{x_1^*x_2\},
\end{equation}
which is equivalent to treating $\mathbb{C}$ as $\mathbb{R}^2$ with the canonical inner product. Then we are able to denote the complex circle as
\begin{equation}
\mathcal{M}_{cc}=\{x\in\mathbb{C}:x^*x=1\}.
\end{equation}
For a given point $x$ on the manifold $\mathcal{M}_{cc}$, the directions along which it can move are characterized by the tangent vectors. Then the tangent space at the point $x\in\mathcal{M}_{cc}$ can be represented by
\begin{equation}
T_x\mathcal{M}_{cc}=\{z\in\mathbb{C}:z^*x+x^*z=2\left\langle {x,z} \right\rangle=0\}.
\end{equation}

Note that the vector $\mathbf{x}=\mathrm{vec}(\mathbf{F}_\mathrm{RF})$ forms a complex circle manifold $\mathcal{M}_{cc}^m=\{{\mathbf{x}\in\mathbb{C}^m:|\mathbf{x}_1|=|\mathbf{x}_2|=\cdot\cdot\cdot=|\mathbf{x}_m|=1}\}$, where $m=N_tN_\mathrm{RF}^t$. Therefore, the search space of the optimization problem \eqref{analog} is over a product of $m$ circles in the complex plane, which is a Riemannian submanifold of $\mathbb{C}^m$ with the product geometry. Hence, the tangent space at a given point $\mathbf{x}\in\mathcal{M}_{cc}^m$ can be expressed as
\begin{equation}
T_\mathbf{x}\mathcal{M}_{cc}^m=\left\{\mathbf{z}\in\mathbb{C}^m:\Re\left\{\mathbf{z}\circ\mathbf{x}^*\right\}=\mathbf{0}_m\right\}.
\end{equation}

Among all the tangent vectors, similar to the Euclidean space, one of them that is related to the negative \emph{Riemannian gradient} represents the direction of the greatest decrease of a function. Because the complex circle manifold $\mathcal{M}_{cc}^m$ is a Riemannian submanifold of $\mathbb{C}^m$, the Riemannian gradient at $\mathbf{x}$ is a tangent vector $\mathrm{grad}f(\mathbf{x})$ given by the orthogonal projection of the Euclidean gradient $\nabla f(\mathbf{x})$ onto the tangent space $T_{\mathbf{x}}\mathcal{M}_{cc}^m$ \cite{absil2009optimization}:
\begin{equation}\label{rgradient}
\begin{split}
\mathrm{grad}f(\mathbf{x})&=\mathrm{Proj}_\mathbf{x}\nabla f(\mathbf{x})\\
&=\nabla f(\mathbf{x})-\Re\{\nabla f(\mathbf{x})\circ \mathbf{x}^*\}\circ\mathbf{x},
\end{split}
\end{equation}
where the Euclidean gradient of the cost function in \eqref{analog} is
\begin{equation}\label{egradient}
\nabla f(\mathbf{x})=-2(\mathbf{F}_{\mathrm{BB}}^*\otimes\mathbf{I}_{N_t})
\left[\mathrm{vec}(\mathbf{F}_{\mathrm{opt}})-
(\mathbf{F}_{\mathrm{BB}}^T\otimes\mathbf{I}_{N_t})\mathbf{x}\right].
\end{equation}
Solving this Euclidean gradient involves some techniques on complex-valued matrix derivatives, the details can be found in \cite{hjorungnes2011complex}.

\emph{Retraction} is another key factor in manifold optimization, which maps a vector from the tangent space onto the manifold itself. It determines the destination on the manifold when moving along a tangent vector. The retraction of a tangent vector $\alpha\mathbf{d}$ at point $\mathbf{x}\in\mathcal{M}_{cc}^m$ can be stated as
\begin{equation}\label{retraction}
\begin{split}
\mathrm{Retr}_\mathbf{x}:&T_{\mathbf{x}}\mathcal{M}_{cc}^m\to\mathcal{M}_{cc}^m:\\
&\alpha\mathbf{d}\mapsto\mathrm{Retr}_\mathbf{x}(\alpha\mathbf{d})=\mathrm{vec}\left[\frac{(\mathbf{x}+\alpha\mathbf{d})_i}{|(\mathbf{x}+\alpha\mathbf{d})_i|}
\right].
\end{split}
\end{equation}

Equipped with the tangent space, Riemannian gradient and retraction of the complex circle manifold $\mathcal{M}_{cc}^m$, a line search based conjugate gradient method \cite{manopt}, which is a classical algorithm in the Euclidean space, can be developed to design the analog precoder as shown in Algorithm 1.
\floatname{algorithm}{Algorithm 1\hspace{-0.5em}}
\begin{algorithm}[h]
\caption{Conjugate Gradient Algorithm for Analog Precoding Based on Manifold Optimization}
\label{conjugategradient}
\begin{algorithmic}[1]
\REQUIRE
$\mathbf{F}_{\mathrm{opt}},\mathbf{F}_{\mathrm{BB}},\mathbf{x}_0\in\mathcal{M}_{cc}^m$
\STATE $\mathbf{d}_0=-\mathrm{grad}f(\mathbf{x}_0)$ and $k=0$;
\REPEAT
\STATE Choose Armijo backtracking line search step size $\alpha_k$;
\STATE Find the next point $\mathbf{x}_{k+1}$ using retraction in \eqref{retraction}: 
$\mathbf{x}_{k+1}=\mathrm{Retr}_{\mathbf{x}_k}(\alpha_k\mathbf{d}_k);$
\STATE Determine Riemannian gradient $\mathbf{g}_{k+1}=\mathrm{grad}f(\mathbf{x}_{k+1})$ according to \eqref{rgradient} and \eqref{egradient};
\STATE Calculate the vector transports $\mathbf{g}_k^+$ and $\mathbf{d}_k^+$ of gradient $\mathbf{g}_k$ and conjugate direction $\mathbf{d}_k$ from $\mathbf{x}_k$ to $\mathbf{x}_{k+1}$;\label{s6}
\STATE Choose Polak-Ribiere parameter $\beta_{k+1}$;\label{s7}
\STATE Compute conjugate direction $\mathbf{d}_{k+1}=-\mathbf{g}_{k+1}+\beta_{k+1} \mathbf{d}_k^+$;\label{s8}
\STATE $k\leftarrow k+1$;
\UNTIL a stopping criterion triggers.
\end{algorithmic}
\end{algorithm}

Algorithm 1 utilizes the well-known Armijo backtracking line search step and Polak-Ribiere parameter to guarantee the objective function to be non-increasing in each iteration \cite{bertsekas1999nonlinear}. In addition, since Steps \ref{s7} and \ref{s8} involve the operations between two vectors in different tangent spaces $T_{\mathbf{x}_k}\mathcal{M}_{cc}^m$ and $T_{\mathbf{x}_{k+1}}\mathcal{M}_{cc}^m$, which cannot be combined directly, a mapping between two tangent vectors in different tangent spaces called \emph{transport} is introduced. The transport of a tangent vector $\mathbf{d}$ from $\mathbf{x}_k$ to $\mathbf{x}_{k+1}$ can be specified as
\begin{equation}
\begin{split}
\mathrm{Transp_{\mathbf{x}_k\to\mathbf{x}_{k+1}}}:&T_{\mathbf{x}_k}\mathcal{M}_{cc}^m\to T_{\mathbf{x}_{k+1}}\mathcal{M}_{cc}^m:\\
&\mathbf{d}\mapsto \mathbf{d}-\Re\{\mathbf{d}\circ \mathbf{x}_{k+1}^*\}\circ\mathbf{x}_{k+1},
\end{split}
\end{equation}
which is accomplished in Step \ref{s6}. According to Theorem 4.3.1 in \cite{absil2009optimization}, Algorithm 1 is guaranteed to converge to a critical point, i.e., the point where the gradient of the objective function is zero.

\subsection{Hybrid Precoder Design}\label{C}
With Algorithm 1 at hand, the hybrid precoder design via alternating minimization for the fully-connected structure is described in the \textbf{MO-AltMin Algorithm} by solving problems \eqref{digital} and \eqref{analog} iteratively. 
To satisfy the power constraint in \eqref{problemformulation}, we normalize $\mathbf{F}_\mathrm{BB}$ by a factor of $\frac{\sqrt{N_s}}{\left\Vert\mathbf{F}_\mathrm{RF}\mathbf{F}_\mathrm{BB}\right\Vert_F}$ at Step \ref{s71}. The following lemma help reveal the effect of this normalization.
\floatname{algorithm}{MO-AltMin Algorithm:}
\begin{algorithm}[h]
\caption{Manifold Optimization Based Hybrid Precoding for the Fully-connected Structure}
\label{alternating}
\begin{algorithmic}[1]
\REQUIRE
$\mathbf{F}_{\mathrm{opt}}$
\STATE Construct $\mathbf{F}_\mathrm{RF}^{(0)}$ with random phases and set $k=0$;
\REPEAT 
\STATE Fix $\mathbf{F}_\mathrm{RF}^{(k)}$, and $\mathbf{F}_\mathrm{BB}^{(k)}=\mathbf{F}_\mathrm{RF}^{(k)\dag}\mathbf{F}_\mathrm{opt}$;\label{s4}
\STATE Optimize $\mathbf{F}_\mathrm{RF}^{(k+1)}$ using Algorithm 1 when $\mathbf{F}_\mathrm{BB}^{(k)}$ is fixed;\label{s3}
\STATE $k\leftarrow k+1$;
\UNTIL a stopping criterion triggers;
\STATE For the digital precoder at the transmit end, normalize
$\widehat{\mathbf{F}}_\mathrm{BB}=\frac{\sqrt{N_s}}{\left\Vert\mathbf{F}_\mathrm{RF}\mathbf{F}_\mathrm{BB}\right\Vert_F}\mathbf{F}_\mathrm{BB}.$\label{s71}
\end{algorithmic}
\end{algorithm}
\begin{lem}\label{lem1}
If the Euclidean distance before normalization is $\left\Vert \mathbf{F}_\mathrm{opt}-\mathbf{F}_\mathrm{RF}\mathbf{F}_\mathrm{BB}\right\Vert _F\le\delta$, then after normalization we have $\left\Vert \mathbf{F}_\mathrm{opt}-\mathbf{F}_\mathrm{RF}\widehat{\mathbf{F}}_\mathrm{BB}\right\Vert _F\le 2\delta$.
\end{lem}

\begin{IEEEproof}
Define the normalization factor $\frac{\sqrt{N_s}}{\left\Vert\mathbf{F}_\mathrm{RF}\mathbf{F}_\mathrm{BB}\right\Vert_F}=\frac{1}{\lambda}$ and thus $\left\Vert \mathbf{F}_\mathrm{RF}\mathbf{F}_\mathrm{BB}\right\Vert _F=\lambda\sqrt{N_s}=\lambda\left\Vert \mathbf{F}_\mathrm{opt}\right\Vert _F$.

By norm inequality, we have
\begin{equation}
\begin{split}
\left\|\mathbf{F_{\mathrm{opt}}-F_{\mathrm{RF}}F_{\mathrm{BB}}}\right\|_F&\ge|\left\|\mathbf{F_{\mathrm{opt}}}\right\|_F-\left\|\mathbf{F_{\mathrm{RF}}F_{\mathrm{BB}}}\right\|_F|\\
&=|1-\lambda|\left\|\mathbf{F_{\mathrm{opt}}}\right\|_F,
\end{split}
\end{equation}
which is equivalent to $\left\|\mathbf{F_{\mathrm{opt}}}\right\|_F\le\frac{1}{|\lambda-1|}\delta$.

When $\lambda\ne1$, which indicates $\left\|\mathbf{F_{\mathrm{opt}}-F_{\mathrm{RF}}F_{\mathrm{BB}}}\right\|_F\ne0$,
\begin{equation}
\begin{split}
&\quad\,\left\|\mathbf{F_{\mathrm{opt}}-F_{\mathrm{RF}}}{\widehat{\mathbf{F}}_\mathrm{BB}}\right\|_F\\
&=\left\|\mathbf{F_{\mathrm{opt}}-F_{\mathrm{RF}}F_{\mathrm{BB}}}+\left(1-\frac{1}{\lambda}\right)\mathbf{F_{\mathrm{RF}}F_{\mathrm{BB}}}\right\|_F\\
&\le\left\|\mathbf{F_{\mathrm{opt}}-F_{\mathrm{RF}}F_{\mathrm{BB}}}\right\|_F+\left|1-\frac{1}{\lambda}\right|\left\|\mathbf{F_{\mathrm{RF}}F_{\mathrm{BB}}}\right\|_F\\
&\le\delta+\left|\lambda-1\right|\left\|\mathbf{F_{\mathrm{opt}}}\right\|_F\le\delta+\frac{|\lambda-1|}{|\lambda-1|}\delta= 2\delta.
\end{split}
\end{equation}
\end{IEEEproof}

Lemma \ref{lem1} shows that as long as we can make the Euclidean distance between the optimal digital precoder and the hybrid precoders sufficiently small when ignoring the power constraint in \eqref{problemformulation}, the normalization step will also achieve a small distance to the optimal digital precoder.

Since the objective function in problem \eqref{problemformulation} is minimized at Steps \ref{s4} and \ref{s3}, each iteration will never increase it. In addition, the objective function is non-negative. These two properties together guarantee that the MO-AltMin algorithm can converge to a feasible solution. Although the optimality of alternating minimization algorithms for general non-convex problems is still an open problem \cite{grippo2000convergence}, simulation results in Section \ref{simulation} will show that the proposed algorithm can provide near-optimal performance.

However, the complexity of the MO-AltMin algorithm is relatively high. In each iteration, the update of the analog precoder involves a line search algorithm, i.e., Algorithm 1, so the nested loops in the MO-AltMin algorithm will slow down the whole solving procedure. Furthermore, the Kronecker products in \eqref{egradient} will result in two matrices of dimension $N_\mathrm{RF}^tN_t\times N_sN_t$, which scales with the antenna size and results in an exponential increase of the computational complexity in the MO-AltMin algorithm.
Despite the high complexity, we note that the MO-AltMin algorithm based on manifold optimization directly solves the hybrid precoder design problem \eqref{problemformulation} under unit modulus constraints, which will improve the spectral efficiency when compared to existing algorithms. Therefore, this algorithm can serve as a benchmark of the performance in terms of spectral efficiency, and we will seek a low-complexity algorithm in the next section.

\section{Low-Complexity Hybrid Precoding for the Fully-connected Structure}\label{IV}
Although the MO-AltMin algorithm can directly handle the unit modulus constraints, the number of such constraints may be substantially large due to the large-size antenna array. Thus, the high computational complexity will prevent its practical implementation. It motivates us to develop a hybrid precoding algorithm with lower computational complexity and slight performance loss. In this section, by utilizing the orthogonal property of the digital precoder, we will propose a low-complexity design for the analog precoder subject to unit modulus constraints. Thanks to the orthogonal property of the digital precoder, the phases of the analog precoder can be extracted from the phases of an equivalent precoder determined by the digital precoder and the unconstrained optimal digital precoder. Though it will incur some performance loss compared to the manifold based algorithm, simulations will demonstrate its performance gains over existing algorithms.

\subsection{Digital Baseband Precoder Structure}
Note that the columns of the unconstrained optimal precoding matrix $\Fopt$ are mutually orthogonal in order to mitigate the interference between the multiplexed streams. Inspired by this structure of the unconstrained precoding solution, we impose a similar constraint that the columns of the digital precoding matrix should be mutually orthogonal, i.e.,
\begin{equation}
\FBB^H\FBB=\alpha\FDD^H\alpha\FDD=\alpha^2\mathbf{I}_{N_s},
\end{equation}
where $\FDD$ is a unitary matrix with the same dimension as $\FBB$. 
Although there is no existing conclusion on the optimal structure of the digital precoder in hybrid precoding, it is natural and intriguing to investigate the hybrid precoder design under such an orthogonal constraint of the digital precoder.
More importantly, this orthogonal constraint creates the potential for the analog precoder $\FRF$ to get rid of the product form with $\FBB$, which will help significantly simplify the analog precoder design.

\subsection{Hybrid precoder design}\label{issue}
By replacing $\FBB$ with $\alpha \FDD$, the objective function in \eqref{problemformulation} can be further recast as
\begin{equation}\label{newobj}
\begin{split}
&\relphantom{=}\left\Vert\Fopt-\FRF\FBB\right\Vert_F^2\\
&=\Tr\left(\Fopt^H\Fopt\right)-\Tr\left(\Fopt^H\FRF\FBB\right)\\
&\relphantom{=}-\Tr\left(\FBB^H\FRF^H\Fopt\right)+\Tr\left(\FBB^H\FRF^H\FRF\FBB\right)\\
&=\left\Vert\Fopt\right\Vert_F^2-2\alpha\Re\Tr\left(\FDD\Fopt^H\FRF\right)\\
&\relphantom{=}+\alpha^2\left\Vert\FRF\FDD\right\Vert_F^2.
\end{split}
\end{equation}
Obviously, when $\alpha=\frac{\Re\Tr\left(\FDD\Fopt^H\FRF\right)}{\left\Vert\FRF\FDD\right\Vert_F^2}$, the objective function $\left\Vert\Fopt-\FRF\FBB\right\Vert_F^2$ in \eqref{newobj} has the minimum value, given by $\Vert\Fopt\Vert_F^2-\frac{\left\{\Re\Tr\left(\FDD\Fopt^H\FRF\right)\right\}^2}{\left\Vert\FRF\FDD\right\Vert_F^2}$.
Note that the square of the Frobenius norm $\left\Vert\FRF\FDD\right\Vert_F^2$ has the following upper bound
\begin{equation}\label{upper}
\begin{split}
\left\Vert\FRF\FDD\right\Vert_F^2&=
\Tr\left(\FDD^H\FRF^H\FRF\FDD\right)\\
&=\Tr\left\{ {\left( {\begin{array}{*{20}{c}}
{{{\mathbf{I}}_{{N_s}}}}&{}\\
{}&{\mathbf{0}}
\end{array}} \right){{\mathbf{K}^H}}{{\FRF^H}}{\FRF\mathbf{K}}} \right\}\\
&\le\Tr\left\{ {{{\mathbf{K}}^H}{\FRF^H}}{\FRF\mathbf{K}} \right\}\\
&=\left\Vert\FRF\right\Vert_F^2,
\end{split}
\end{equation}
where $\FDD\FDD^H=\mathbf{K}\left( {\begin{array}{*{20}{c}}
{{{\mathbf{I}}_{{N_s}}}}&{}\\
{}&{\mathbf{0}}
\end{array}} \right)\mathbf{K}^H$ is the SVD of $\FDD\FDD^H$ and the equality holds when $N_\mathrm{RF}^t=N_s$, i.e., $\FDD$ is a square matrix. Hence, the objective function in \eqref{problemformulation} is upper bounded by $\Vert\Fopt\Vert_F^2-\frac{\left\{\Re\Tr\left(\FDD\Fopt^H\FRF\right)\right\}^2}{\left\Vert\FRF\right\Vert_F^2}$. In order to make $\FRF$ get rid of the product with $\FBB$, we choose to add the constant term $\left(\frac{1}{2\left\Vert\FRF\right\Vert_F^2}-1\right)\left\Vert\Fopt\right\Vert_F^2+\frac{1}{2}$ to the bound and multiply it by the positive constant term $2\left\Vert\FRF\right\Vert_F^2$. Then we have
\begin{equation}\label{transfer}
\begin{split}
&\relphantom{=}\left\Vert\Fopt\right\Vert_F^2-2\Re\Tr\left(\FDD\Fopt^H\FRF\right)+\left\Vert\FRF\right\Vert_F^2\\
&=\Tr\left(\FRF^H\FRF\right)-2\Re\Tr\left(\FDD\Fopt^H\FRF\right)\\
&\relphantom{=}+\Tr\left(\FDD\Fopt^H\Fopt\FDD^H\right)\\
&=\left\Vert\Fopt\FDD^H-\FRF\right\Vert_F^2.
\end{split}
\end{equation}

Since directly optimizing the objective function \eqref{newobj} will still incur high complexity, we intend to adopt the upper bound \eqref{transfer} as the objective function rather than the original one.
In addition, as we can satisfy the transmit power constraint by normalization after updating the hybrid precoders alternately, which has been shown in the MO-AltMin algorithm and Lemma \ref{lem1}, here we also temporarily remove the power constraint. Thus, by adopting \eqref{transfer} as the objective function, the hybrid precoder design problem is given as
\begin{equation}\label{new}
\begin{aligned}
&\underset{\mathbf{F}_\mathrm{RF},\mathbf{F}_\mathrm{DD}}{\mathrm{minimize}} && \left\Vert\Fopt\FDD^H-\FRF\right\Vert_F^2\\
&\mathrm{subject\thinspace to}&&
\begin{cases}
|{(\mathbf{F}_\mathrm{RF}})_{i,j}|=1, \forall i,j\\
\FDD^H\FDD=\mathbf{I}_{N_s}.\\
\end{cases}
\end{aligned}
\end{equation}

The problem formulation \eqref{new} implies that we only need to seek a unitary precoding matrix $\FDD$, and then a corresponding precoding matrix $\FBB$ with orthogonal columns can be obtained. When applying alternating minimization, it turns out that the objective function in \eqref{new} significantly simplifies the analog precoder design. More specifically, since the matrix $\FRF$ gets rid of the product form with $\FBB$, it has a closed-form solution
\begin{equation}
\arg\left(\mathbf{F}_\mathrm{RF}\right)=\arg\left(\mathbf{F}_{\mathrm{opt}}\FDD^H\right),
\end{equation}
where $\mathrm{arg}(\mathbf{A})$ generates a matrix containing the phases of the entries of $\mathbf{A}$. Thus, it shows that the phases of $\FRF$ can be extracted from the phases of an equivalent precoder $\Fopt\FDD^H$. This closed-form solution can also be viewed as the Euclidean projection of $\Fopt\FDD^H$ on the feasible set $\mathcal{A}_f$ of the analog precoder.

For the digital precoder design, regarding $\mathbf{F}_\mathrm{RF}$ as fixed, we try to solve a digital precoder which optimizes the following problem
\begin{equation}
\begin{aligned}\label{p25}
&\underset{\mathbf{F}_\mathrm{DD}}{\mathrm{minimize}} && \left\Vert\Fopt\FDD^H-\FRF\right\Vert_F^2\\
&\mathrm{subject\thinspace to}&&
\FDD^H\FDD=\mathbf{I}_{N_s}.\\
\end{aligned}
\end{equation}
Since problem \eqref{p25} only has one optimization variable $\FDD$, it is equivalent to
\begin{equation}
\begin{aligned}
&\underset{\mathbf{F}_\mathrm{DD}}{\mathrm{maximize}} && \Re\Tr\left(\FDD\Fopt^H\FRF\right)\\
&\mathrm{subject\thinspace to}&&
\FDD^H\FDD=\mathbf{I}_{N_s}.\\
\end{aligned}
\end{equation}
According to the definition of the dual norm, we have
\begin{equation}
\begin{split}
\Re\Tr\left(\FDD\Fopt^H\FRF\right)&\le\left|\Tr\left(\FDD\Fopt^H\FRF\right)\right|\\
&\overset{(a)}{\le}\left\Vert\FDD^H\right\Vert_\infty\cdot\left\Vert\Fopt^H\FRF\right\Vert_1\\
&=\left\Vert\Fopt^H\FRF\right\Vert_1\\
&=\sum_{i=1}^{N_s}{\sigma_i},
\end{split}
\end{equation}
where (a) follows the H\"{o}lder's inequality, $\left\Vert\cdot\right\Vert_\infty$ and $\left\Vert\cdot\right\Vert_1$ stand for the infinite and one Schatten norms \cite{horn2012matrix}. The equality is established only when
\begin{equation}\label{uv}
\FDD=\mathbf{V}_1\mathbf{U}^H,
\end{equation}
where $\Fopt^H\FRF=\mathbf{U\Sigma V}^H=\mathbf{U}\mathbf{SV}_1^H$, which is the SVD of $\Fopt^H\FRF$, and $\mathbf{S}$ is a diagonal matrix whose elements are the first $N_s$ nonzero singular values $\sigma_1,\cdots,\sigma_{N_s}$.
This result bears some similarity to the solution of the orthogonal Procrustes problem (OPP) \cite{gower2004procrustes}, although the formulation is slightly different\footnote{OPP tries to minimize $\left\Vert\mathbf{A\Omega-B}\right\Vert_F$, where the optimization variable $\mathbf{\Omega}$ is a square unitary matrix.}.
\floatname{algorithm}{PE-AltMin Algorithm:}
\begin{algorithm}[h]
\caption{A Low-Complexity Algorithm for the Fully-connected Structure}
\begin{algorithmic}[1]\label{lowcom}
\REQUIRE
$\Fopt$
\STATE Construct $\FRF^{(0)}$ with random phases and set $k=0$;
\REPEAT 
\STATE Fix $\FRF^{(k)}$, compute the SVD: $\Fopt^H\FRF^{(k)}=\mathbf{U}^{(k)}\mathbf{S}^{(k)}{\mathbf{V}_1^{(k)}}^H$; 
\STATE $\FDD^{(k)}=\mathbf{V}_1^{(k)}{\mathbf{U}^{(k)}}^H$;
\STATE Fix $\FDD^{(k)}$, and 
$\arg\left\{\FRF^{(k+1)}\right\}=\arg\left({\Fopt\FDD^{(k)}}^H\right)$;\label{step5}
\STATE $k\leftarrow k+1$;
\UNTIL a stopping criterion triggers;
\STATE For the digital precoder at the transmit end, normalize
$\widehat{\mathbf{F}}_\mathrm{BB}=\frac{\sqrt{N_s}}{\left\Vert\FRF\FDD\right\Vert_F}\FDD.$
\end{algorithmic}
\end{algorithm}

Based on the two closed-form solutions for the analog and digital precoders, we summarize our new design as the \textbf{PE-AltMin Algorithm}. There are several issues involved in the PE-AltMin algorithm that require some further remarks.


\emph{(1) Complexity:} In both the MO-AltMin and PE-AltMin algorithms, the updating rules of the digital precoders are given by closed-form solutions and thus these two algorithms are of comparable complexity in the digital parts. Furthermore, in the hybrid precoding system, the dimension of the analog precoder is much higher than that of the digital precoder, which makes the complexity of the algorithms predominated by the analog part.

In each iteration of the MO-AltMin algorithm, a conjugate gradient descent search is needed to update the analog precoder. In particular, when updating the analog precoder in each iteration, we need to search on the complex circle manifold repeatedly to find a local optimum with zero gradient of the cost function. Additionally, the computation of the large-size matrices, i.e., matrices of dimension $N_\mathrm{RF}^tN_t\times N_sN_t$, will be involved in the gradient descent procedure due to the use of the Kronecker products. More importantly, the conjugate gradient descent, which is an iterative procedure itself, is nested into each alternating minimization iteration. This nested iteration structure will dramatically degrade the computational efficiency of the MO-AltMin algorithm.
On the contrary, in each iteration of the PE-AltMin algorithm, the update of the analog precoder is simply realized by a phase extraction operation of the matrix $\Fopt\FDD^H$, whose dimension is $N_t\times N_\mathrm{RF}^t$. Compared with the MO-AltMin algorithm, it is safe to conclude that the PE-AltMin algorithm is with a much lower complexity, which is also numerically observed in our simulations.

\emph{(2) Accuracy of the approximation:} In the formulation of \eqref{new}, we try to minimize an upper bound given by \eqref{upper} rather than directly minimizing the original objective function. Therefore, the effectiveness of this strategy depends on how tight the upper bound is when $N_s\le N_\mathrm{RF}^t<2N_s$. According to \eqref{upper}, we can quantify the gap between $\Vert\FRF\FDD\Vert_F^2$ and $\Vert\FRF\Vert_F^2$ as $\Vert\FRF\mathbf{K}_1\Vert_F^2$, where $\mathbf{K}_1$ is comprised of the rightmost $N_\mathrm{RF}^t-N_s$ columns of $\mathbf{K}$. Note that when $N_\mathrm{RF}^t=N_s$, the upper bound is tight, i.e., the equality holds in \eqref{upper}. Furthermore, when increasing $N_\mathrm{RF}^t$ from $N_s$ to $2N_s-1$, the gap $\Vert\FRF\mathbf{K}_1\Vert_F^2$ will become larger with increasing $N_\mathrm{RF}^t$, which will result in some performance loss. The impact of adopting this upper bound as the objective will be shown in Section \ref{simulation} via simulations.

\emph{(3) Calculation of $\alpha$:} Although we have stated that a value of $\alpha$ can be found to construct a matrix $\FBB$ corresponding to each $\FDD$, we do not need to actually calculate $\alpha$ in the PE-AltMin algorithm for the following reasons. At the transmitter side, even though we compute $\FBB=\alpha\FDD$ in the last step of the PE-AltMin algorithm, this digital precoder should be immediately normalized. Hence, the overall procedure is equivalent to directly normalizing $\FDD$ to satisfy the power constraint without knowing $\alpha$. At the receiver side, we note that the spectral efficiency \eqref{spectralefficiency} will not be influenced by the constant factor $\alpha$ multiplied with $\mathbf{W}_\mathrm{BB}$. That is because the decoder $\mathbf{W}_\mathrm{BB}$ takes effects on both received signals and noise, and thus signal-to-noise ratio (SNR) will not change due to the constant factor $\alpha$. Moreover, the avoidance of calculating $\alpha$ will further reduce the complexity of the PE-AltMin algorithm.

\section{Hybrid Precoding for the Partially-connected Structure}
Different from the fully-connected structure, the partially-connected structure shown in Fig. \ref{subarray} \cite{singh2014feasibility,zhang2015massive,el2013multimode}, also called the \emph{array of subarray structure}, employs notably less phase shifters and is advocated for energy-efficient mmWave MIMO systems \cite{7248508,han2015large}. Particularly, the output signal of each RF chain is only connected with $N_t/N_\mathrm{RF}^t$ antennas, which reduces the hardware complexity in the RF domain. Therefore, the analog precoder $\mathbf{F}_\mathrm{RF}$ in this structure belongs to a set of block matrices $\mathcal{A}_p$, where each block is an $N_t/N_\mathrm{RF}^t$ dimension vector with unit modulus elements and the structure of $\FRF$ can be depicted as
\begin{equation}
\FRF=\left[ {\begin{array}{*{20}{c}}
{\mathbf{p}_1}&{\mathbf{0}}&{\cdots}&{\mathbf{0}}\\
{\mathbf{0}}&{\mathbf{p}_2}&{}&{\mathbf{0}}\\
{\vdots}&{}&{\ddots}&{\vdots}\\
{\mathbf{0}}&{\mathbf{0}}&{\cdots}&{\mathbf{p}_{N_\mathrm{RF}^t}}
\end{array}} \right]_{},
\end{equation}
where $\mathbf{p}_i=\left[\exp\left(j\theta_{(i-1)\frac{N_t}{N_\mathrm{RF}^t}+1}\right),\cdots,\exp\left(j\theta_{i\frac{N_t}{N_\mathrm{RF}^t}}\right)\right]^T$ and $\theta_i$ stands for the phase of the $i$th phase shifter. In this section, we will propose an AltMin algorithm for this structure. Surprisingly, optimal solutions can be found for both subproblems of analog and digital precoders.
\subsection{Analog RF Precoder Design}
Due to the special structure of the constraint on $\mathbf{F}_\mathrm{RF}$, in the product $\mathbf{F}_\mathrm{RF}\mathbf{F}_\mathrm{BB}$, each nonzero element of $\mathbf{F}_\mathrm{RF}$ is multiplied by a corresponding row extracted from $\mathbf{F}_\mathrm{BB}$. Thus, the power constraint in \eqref{problemformulation} at the transmit side can be recast as
\begin{equation}\label{powerconstraintsimplify}
\left\Vert\FRF\FBB\right\Vert_F^2=\frac{N_t}{N_\mathrm{RF}^t}\left\Vert\FBB\right\Vert_F^2=N_s.
\end{equation}
Therefore, the analog precoder design is formulated as
\begin{equation}\label{partial-analog}
\begin{aligned}
&\underset{\mathbf{F}_\mathrm{RF}}{\mathrm{minimize}} && \left\Vert \mathbf{F}_\mathrm{opt}-\mathbf{F}_\mathrm{RF}\mathbf{F}_\mathrm{BB}\right\Vert _F^2\\
&\mathrm{subject\thinspace to} &&{\mathbf{F}_\mathrm{RF}}\in\mathcal{A}_p.
\end{aligned}
\end{equation}
Also, due to the same property of $\FRF$, problem \eqref{partial-analog} can be reformulated as
\begin{equation}
\begin{aligned}
&\underset{\{\theta_i\}_{i=1}^{N_t}}{\mathrm{minimize}} && \left\Vert (\Fopt)_{i,:}-e^{j\theta_i}(\FBB)_{l,:}\right\Vert _2^2,
\end{aligned}
\end{equation} 
where $l=\left\lceil {i\frac{N_\mathrm{RF}^t}{N_t}} \right\rceil$. This is basically a vector approximation problem using phase rotation, and there exists a closed-form expression for nonzero elements in $\mathbf{F}_\mathrm{RF}$, given by
\begin{equation}\label{analogclosed}
\begin{split}
\arg\left\{(\mathbf{F}_\mathrm{RF})_{i,l}\right\}=\arg\left\{(\mathbf{F}_{\mathrm{opt}})_{i,:}{(\mathbf{F}_{\mathrm{BB}})_{l,:}}^H\right\},\\
1\le i\le N_t,l=\left\lceil {i\frac{N_\mathrm{RF}^t}{N_t}} \right\rceil.
\end{split}
\end{equation}
We note that the special characteristic of $\FRF$ simplifies the analog precoder design and makes the unit modulus constraint no longer an intractable issue in the partially-connected structure.
\subsection{Digital Baseband Precoder Design}
According to \eqref{powerconstraintsimplify}, the precoder design at the transmit side can be rewritten as the following problem
\begin{equation}\label{QCQP}
\begin{aligned}
&\underset{\mathbf{F}_\mathrm{BB}}{\mathrm{minimize}} && \left\Vert \mathbf{F}_\mathrm{opt}-\mathbf{F}_\mathrm{RF}\mathbf{F}_\mathrm{BB}\right\Vert _F^2\\
&\mathrm{subject\thinspace to} &&{\left\| \bf{F_{\rm{BB}}} \right\|}_F^2=\frac{N_\mathrm{RF}^tN_s}{N_t}.
\end{aligned}
\end{equation}
Problem \eqref{QCQP} is a non-convex quadratic constraint quadratic programming (QCQP) problem, which can be reformulated as a homogeneous QCQP problem:
\begin{equation}\label{homo}
\begin{aligned}
&\underset{\mathbf{Y}\in\mathbb{H}^n}{\mathrm{minimize}} && \mathrm{Tr}(\mathbf{CY})\\
&\mathrm{subject\thinspace to}&&
\begin{cases}
\mathrm{Tr}(\mathbf{A}_1\mathbf{Y})=\frac{N_\mathrm{RF}^tN_s}{N_t}\\
\mathrm{Tr}(\mathbf{A}_2\mathbf{Y})=1\\
\mathbf{Y}\succeq 0,\;\mathrm{rank}(\mathbf{Y})=1,
\end{cases}
\end{aligned}
\end{equation}
with $\mathbb{H}^n$ being the set of $n=N_\mathrm{RF}^tN_s+1$ dimension complex Hermitian matrices. In addition, $\mathbf{y}=\left[{\begin{array}{*{20}{c}}{\mathrm{vec}(\mathbf{F}_\mathrm{BB})}&{t}\end{array}} \right]^T$ with an auxiliary variable $t$, $\mathbf{Y}=\mathbf{yy}^H$, $\mathbf{f}=\mathrm{vec}(\mathbf{F}_\mathrm{opt})$, and
\begin{equation*}
\mathbf{A}_1=\left[ {\begin{array}{*{20}{c}}
{\mathbf{I}_{n-1}}&{\mathbf{0}}\\
{\mathbf{0}}&{0}
\end{array}} \right],
\mathbf{A}_2=\left[ {\begin{array}{*{20}{c}}
{\mathbf{0}_{n-1}}&{\mathbf{0}}\\
{\mathbf{0}}&{1}
\end{array}} \right],
\end{equation*}
\begin{equation*}
\mathbf{C}=\left[ {\begin{array}{*{20}{c}}
{(\mathbf{I}_{N_s}\otimes\mathbf{F}_{\mathrm{RF}})^H(\mathbf{I}_{N_s}\otimes\mathbf{F}_{\mathrm{RF}})}&
{-(\mathbf{I}_{N_s}\otimes\mathbf{F}_{\mathrm{RF}})^H\mathbf{f}}\\
{-\mathbf{f}^H(\mathbf{I}_{N_s}\otimes\mathbf{F}_{\mathrm{RF}})}&
{\mathbf{f}^H\mathbf{f}}
\end{array}} \right].
\end{equation*}
The derivation and the formulation of the homogeneous QCQP problem can be found in Appendix A.

In fact, the most difficult part in problem \eqref{homo} is the rank constraint, which is non-convex with respect to $\mathbf{Y}$. Thus, we first drop it to obtain a relaxed version of \eqref{homo}, i.e., a semidefinite relaxation (SDR) problem as follows.
\begin{equation}\label{homo1}
\begin{aligned}
&\underset{\mathbf{Y}\in\mathbb{H}^n}{\mathrm{minimize}} && \mathrm{Tr}(\mathbf{CY})\\
&\mathrm{subject\thinspace to}&&
\begin{cases}
\mathrm{Tr}(\mathbf{A}_1\mathbf{Y})=\frac{N_\mathrm{RF}^tN_s}{N_t}\\
\mathrm{Tr}(\mathbf{A}_2\mathbf{Y})=1\\
\mathbf{Y}\succeq 0.
\end{cases}
\end{aligned}
\end{equation}
It has been established that the SDR is tight when the number of constraints is less than three for a complex-valued homogeneous QCQP problem \cite{luo2010semidefinite}. Consequently, problem \eqref{homo1} without the rank-one constraint reduces into a semidefinite programming (SDP) problem and it can be solved by standard convex optimization algorithms \cite{boyd2004convex}, from which we can obtain the globally optimal solution of the digital precoder design problem \eqref{QCQP}. Therefore, a step-by-step summary is provided below as the \textbf{SDR-AltMin Algorithm}.
\floatname{algorithm}{SDR-AltMin Algorithm:}
\begin{algorithm}[h]
\caption{SDR Based Hybrid Precoding for the Partially-connected Structure}\label{partially-alg}
\label{alternating}
\begin{algorithmic}[1]
\REQUIRE
$\mathbf{F}_{\mathrm{opt}}$
\STATE Construct $\mathbf{F}_\mathrm{RF}^{(0)}$ with random phases and set $k=0$;
\REPEAT 
\STATE Fix $\FRF^{(k)}$, solving $\FBB^{(k)}$ using SDR \eqref{homo1};
\STATE Fix $\FBB^{(k)}$, and update $\FRF^{(k+1)}$ by \eqref{analogclosed};
\STATE $k\leftarrow k+1$;
\UNTIL a stopping criterion triggers.
\end{algorithmic}
\end{algorithm}

\subsection{Comparison Between Two Hybrid Precoding Structures}
The main difference between two hybrid precoding structures considered in this paper is the number of phase shifters $N_\mathrm{PS}$ in use for given numbers of data streams, RF chains, and antennas.

In terms of spectral efficiency, the fully-connected structure provides more design degrees of freedom (DoFs) in the RF domain and thus will outperform the partially-connected one. However, when taking power consumption into consideration, it is intriguing to know which structure has better energy efficiency. Energy efficiency is defined as the ratio between spectral efficiency and total power consumption
\begin{equation}\label{energy}
\eta=\frac{R}{P_\mathrm{common}+N_\mathrm{RF}^tP_\mathrm{RF}+N_tP_\mathrm{PA}+N_\mathrm{PS}P_\mathrm{PS}},
\end{equation}
where the unit of $\eta$ is bits/Hz/J and $P_\mathrm{common}$ is the common power of the transmitter. $P_\mathrm{RF}$, $P_\mathrm{PS}$, and $P_\mathrm{PA}$ are the power of each RF chain, phase shifter, and power amplifier, respectively. The number of phase shifters $N_\mathrm{PS}$ can be expressed as follows,
\begin{equation}
N_\mathrm{PS}= \left\{ {\begin{array}{*{20}{l}}
N_tN_\mathrm{RF}^t&\text{fully-connected}\\
N_t&\text{partially-connected}
\end{array}} \right..
\end{equation}
The numerical comparison will be provided in the Section \ref{simulation}.

\section{Hybrid Precoding in MmWave MIMO-OFDM Systems}
In previous sections, we designed hybrid precoders for narrowband mmWave systems. On the other hand, the large available bandwidth is one of the unique characteristics of mmWave systems, and therefore the design of the hybrid precoders should be investigated when multicarrier techniques such as OFDM are utilized to overcome the multipath fading. In this section, we will extend the proposed AltMin algorithms to mmWave MIMO-OFDM systems.

In conventional MIMO-OFDM systems with sub-6 GHz carrier frequencies, digital precoding is performed in the frequency domain for every subcarrier, which can also be adopted in mmWave MIMO-OFDM systems. Futhermore, the digital precoding is followed by an inverse fast Fourier transform (IFFT) operation, which combines the signals of all the subcarriers together. However, since the analog precoding is a post-IFFT processing, the signals of all the subcarriers can only share one common analog precoder in mmWave MIMO-OFDM systems \cite{6824962,6884253}. Under this new restriction, the received signal of each subcarrier after the decoding process can then be expressed as
\begin{equation}
\mathbf{y}[k]=\sqrt{\rho}\mathbf{W}_\mathrm{BB}^H[k]\mathbf{W}_\mathrm{RF}^H\mathbf{H}[k]\mathbf{F}_\mathrm{RF}\mathbf{F}_\mathrm{BB}[k]\mathbf{s}
+\mathbf{W}_\mathrm{BB}^H[k]\mathbf{W}_\mathrm{RF}^H\mathbf{n},
\end{equation}
where $k\in[0,K-1]$ is the subcarrier index. $\mathbf{H}[k]$ is the frequency domain channel matrix for the $k$th subcarrier, which is given by \cite{6884253}
\begin{equation}
\mathbf{H}[k]=\gamma\sum_{i=0}^{N_{cl}-1}\sum_{l=1}^{N_{ray}}{\alpha_{il}\mathbf{a}_r(\phi_{il}^r,\theta_{il}^r)\mathbf{a}_t(\phi_{il}^t,\theta_{il}^t)^H}e^{-j2\pi ik/K},
\end{equation}
where $\gamma=\sqrt{\frac{N_tN_r}{N_{cl}N_{ray}}}$ is the normalization factor and $K$ is the total number of subcarriers. Then the hybrid precoder design in mmWave MIMO-OFDM systems can be formulated as \cite{6884253}
\begin{equation}\label{OFDM}
\begin{aligned}
&\underset{\mathbf{F}_\mathrm{RF},\mathbf{F}_\mathrm{BB}[k]}{\mathrm{minimize}} && \sum_{k=0}^{K-1}\left\Vert \mathbf{F}_\mathrm{opt}[k]-\mathbf{F}_\mathrm{RF}\mathbf{F}_\mathrm{BB}[k]\right\Vert _F^2\\
&\mathrm{subject\thinspace to}&&
\begin{cases}
\FRF\in\mathcal{A}\\
\left\|\mathbf{F}_\mathrm{RF}\mathbf{F}_\mathrm{BB}[k]\right\|_F^2=N_s,
\end{cases}
\end{aligned}
\end{equation}
where $\Fopt[k]$ is the optimal digital precoder for the $k$th subcarrier. While it does not directly maximize the spectral efficiency, similar to the narrowband case, as shown in \cite{el2014spatially,6884253}, the objective is a good surrogate and it will make the problem tractable. 

The alternating minimization framework can be adopted to solve problem \eqref{OFDM}. In particular, the digital precoders of all the subcarriers can be updated in a parallel fashion, since we can get rid of the summation in \eqref{OFDM} when optimizing the digital precoder for each subcarrier. Hence, the solutions from \eqref{ls}, \eqref{uv} and \eqref{homo1} still hold for problem \eqref{OFDM}. Next we will focus on the analog precoder design, which is the main difference from narrowband systems. Then the proposed AltMin algorithms can be applied to the hybrid precoding in OFDM systems.

For the MO-AltMin algorithm, based on the principles of manifold optimization, as mentioned in Section \ref{III}, we first derive the Euclidean gradient of the objective function in \eqref{OFDM} as
\begin{equation}
\begin{split}
&\relphantom{=}\nabla f(\mathbf{x})=-2\sum_{k=0}^{K-1}(\mathbf{F}_{\mathrm{BB}}^*[k]\otimes\mathbf{I}_{N_t})\times\\
&\quad\quad\quad\quad\quad\quad\left[\mathrm{vec}(\mathbf{F}_{\mathrm{opt}}[k])-(\mathbf{F}_{\mathrm{BB}}^T[k]\otimes\mathbf{I}_{N_t})\mathbf{x}\right].
\end{split}
\end{equation}
After calculating this Euclidean gradient, we can still use the projection \eqref{rgradient} and the retraction \eqref{retraction} to obtain the Riemannian gradient and the mapped gradient vector on the manifold, which are key elements in Algorithm 1.

For the PE-AltMin algorithm, after we adopt a similar upper bound and manipulations in Section \ref{IV}, the problem formulation for the analog precoder design is given as
\begin{equation}\label{new}
\begin{aligned}
&\underset{\mathbf{F}_\mathrm{RF}}{\mathrm{minimize}} && \sum_{k=0}^{K-1}\left\Vert\Fopt[k]\FDD^H[k]-\FRF\right\Vert_F^2\\
&\mathrm{subject\thinspace to}&&|{(\mathbf{F}_\mathrm{RF}})_{i,j}|=1, \forall i,j,
\end{aligned}
\end{equation}
which has a closed-form solution
\begin{equation}
\arg\left(\mathbf{F}_\mathrm{RF}\right)=\arg\left(\sum_{k=0}^{K-1}\mathbf{F}_{\mathrm{opt}}[k]\FDD^H[k]\right).
\end{equation}
By substituting Step \ref{step5} in the PE-AltMin algorithm, this solution enables the extension of the PE-AltMin algorithm to OFDM systems.

Similar to the previous two AltMin Algorithms, the SDR-AltMin algorithm can also be realized in OFDM systems. We can get the closed-form solution for the analog precoder as
\begin{equation}
\begin{split}
\arg\left\{(\mathbf{F}_\mathrm{RF})_{i,l}\right\}=\arg\left\{\sum_{k=0}^{K-1}(\mathbf{F}_{\mathrm{opt}}[k])_{i,:}{(\mathbf{F}_{\mathrm{BB}}[k])_{l,:}}^H\right\},\\
1\le i\le N_t,l=\left\lceil {i\frac{N_\mathrm{RF}^t}{N_t}} \right\rceil,
\end{split}
\end{equation}
which can be added in Step 4 in the SDR-AltMin algorithm.

The aforementioned solutions demonstrate that the proposed AltMin algorithms can be directly extended to mmWave MIMO-OFDM systems, and the performance of the proposed AltMin algorithms will be evaluated in the next section.

\section{Simulation Results}\label{simulation}
In this section, we will numerically evaluate the performance of our proposed algorithms. Data streams are sent from a transmitter with $N_t=144$ to a receiver with $N_r=36$ antennas, while both are equipped with USPA. The channel parameters are set as $N_{cl}=5$ clusters, $N_{ray}=10$ rays and the average power of each cluster is $\sigma_{\alpha,i}^2=1$. The azimuth and elevation angles of departure and arrival (AoDs and AoAs) follow the Laplacian distribution with uniformly distributed mean angles and angular spread of 10 degrees. The antenna elements in the USPA are separated by a half wavelength distance and all simulation results are averaged over 1000 channel realizations.
For all the proposed AltMin algorithms, the initial phases of the analog precoder $\FRF$ follow a uniform distribution over $[0,2\pi)$.

\subsection{Spectral Efficiency Evaluation}
Firstly, we investigate the spectral efficiency achieved by different algorithms when the number of RF chains is equal to that of the data streams, i.e., $N_\mathrm{RF}^t=N_\mathrm{RF}^r=N_s$.
\begin{figure}[htbp]
\centering\includegraphics[width=8cm]{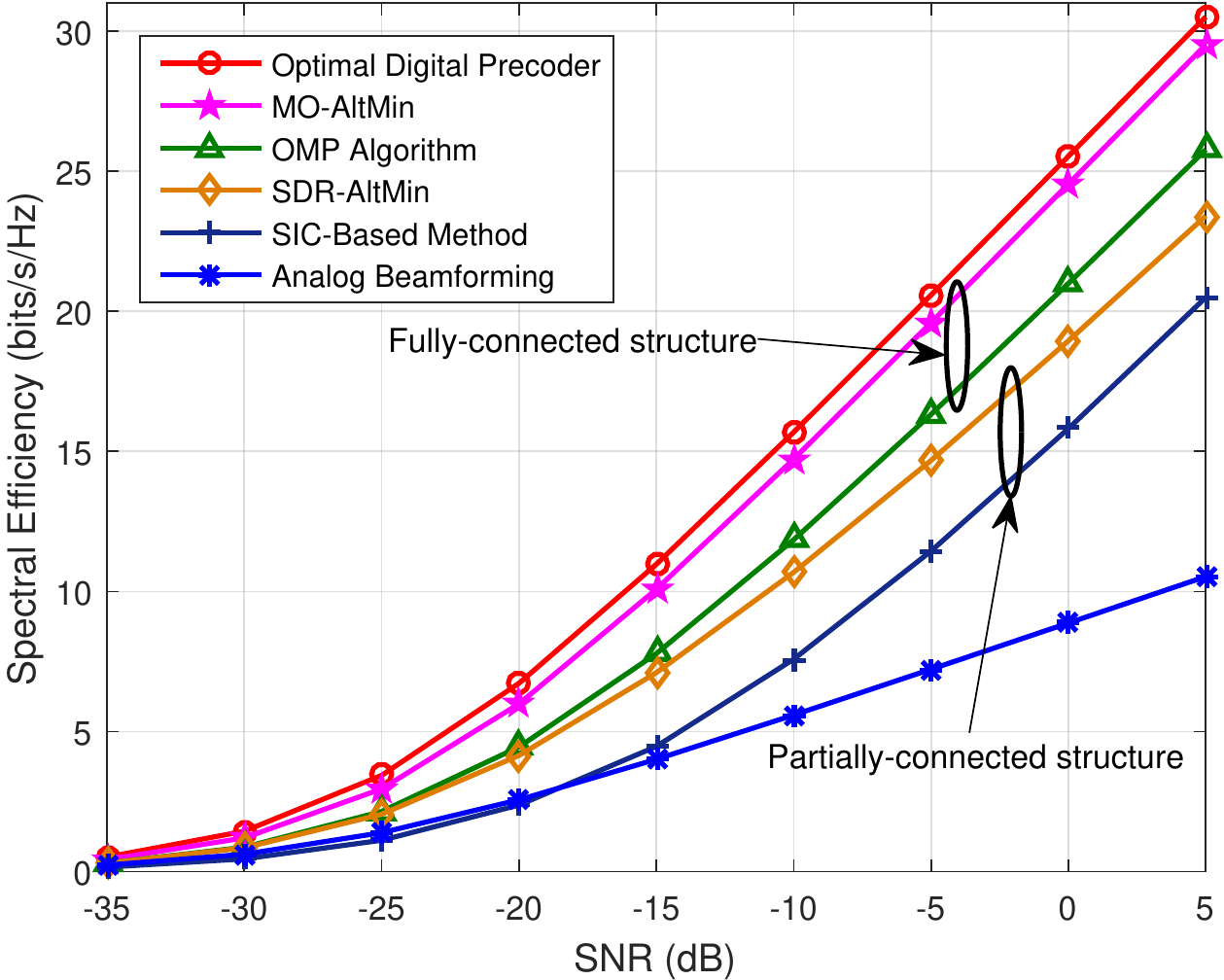}
\caption{Spectral efficiency achieved by different precoding algorithms when $N_\mathrm{RF}^t=N_\mathrm{RF}^r=N_s=3$.}\label{f1}
\end{figure}
This is the worst case since the number of RF chains cannot be smaller under the assumptions in Section \ref{IIA}.
In this case, as shown in Fig. \ref{f1}, for the fully-connected structure, the existing OMP algorithm \cite{el2014spatially} achieve significantly lower spectral efficiency than the optimal digital precoder.
On the contrary, our proposed alternating minimization algorithm\footnote{When optimizing the analog precoder, Algorithm 1 converges within 20 iterations for almost all the channel realizations.}, i.e., the MO-AltMin algorithm, achieves near-optimal performance over the whole SNR range in consideration.
This means that the proposed algorithm can more accurately approximate the optimal digital precoder than existing algorithms, even though the RF chains are limited. In this scenario, the partially-connected structure with the proposed SDR-AltMin algorithm provides substantial performance gains over analog beamforming, especially at high SNRs.
For comparison\footnote{Since the SIC-Based method in \cite{7248508} can only design hybrid precoders at the transmitter, we assume that the optimal digital decoder is adopted at the receiver. Furthermore, the same setting is also employed for the SDR-AltMin algorithm for fair comparison in Fig. \ref{f1}. Nevertheless, in the remainder of this section, both hybrid precoder and decoder are assumed at the transmitter and receiver for the partially-connected structure.}, the SIC-Based method proposed in \cite{7248508} is adopted as a benchmark for the partially-connected structure. Fig. \ref{f1} shows that the SDR-AltMin algorithm outperforms the benchmark. This is mainly because the SDR-AltMin algorithm takes full use of the digital precoder, while the SIC-Based method only uses the digital precoder to allocate power to the data streams under the extra diagonal constraint.

It has been shown in \cite{zhang2014achieving} that when $N_\mathrm{RF}^t\ge 2N_s$ and $N_\mathrm{RF}^r\ge2N_s$, there exists a closed-form solution to the design problem of the fully-connected hybrid precoding, which leads to the same spectral efficiency provided by the optimal digital precoding. Although the hybrid precoder design in this paper aims at cases of $N_s\le N_\mathrm{RF}^t<2N_s$, it is interesting to examine if our proposed algorithm can achieve the comparable performance as the case considered in \cite{zhang2014achieving}.
\begin{figure}[htbp]
\centering\includegraphics[width=8cm]{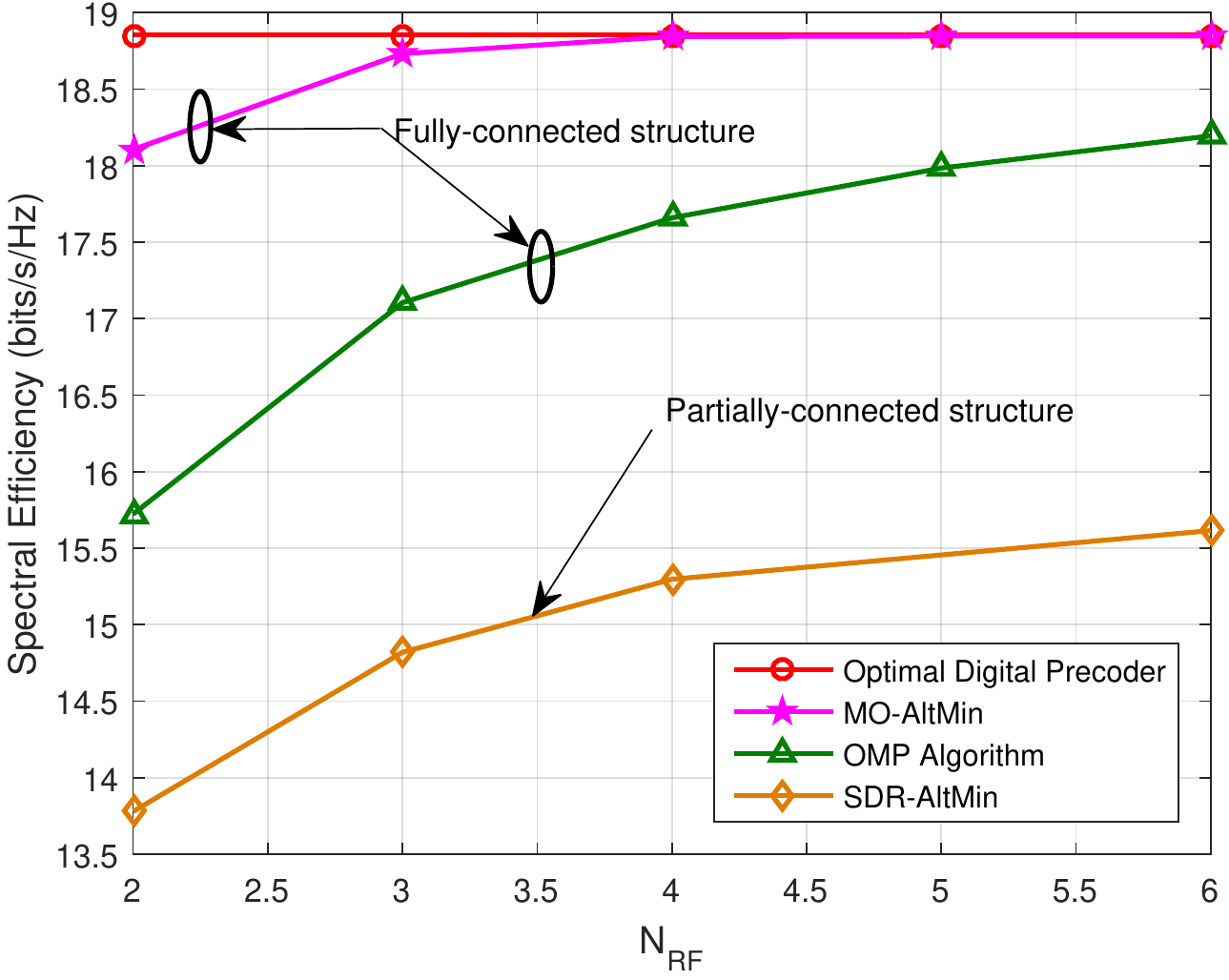}
\caption{Spectral efficiency achieved by different precoding algorithms given $N_s=2$, $N_\mathrm{RF}^t=N_\mathrm{RF}^r=N_\mathrm{RF}$ and $\mathrm{SNR}=0$ dB.}\label{f2}
\end{figure}
Fig. \ref{f2} compares the performance of different precoding schemes for different $N_\mathrm{RF}$.
We see that the proposed MO-AltMin algorithm for the fully-connected structure starts to coincide with the optimal digital precoding when $N_\mathrm{RF}^t=N_\mathrm{RF}^r\ge4$. This result demonstrates that when $N_\mathrm{RF}^t\ge 2N_s$ and $N_\mathrm{RF}^r\ge2N_s$, our proposed algorithm can actually achieve the optimal spectral efficiency, which, however, cannot be achieved by the OMP algorithm. Furthermore, the comparison between two hybrid precoding structures shows that the partially-connected structure, using less phase shifters, does entail some non-negligible performance loss when compared with the fully-connected structure.

\subsection{Energy Efficiency Evaluation}
In this part, we will compare the performance of the two hybrid precoding structures in terms of energy efficiency, as defined in \eqref{energy}. The simulation parameters are set as follows: $P_\mathrm{common}=10$ W, $P_\mathrm{RF}=100$ mW, $P_\mathrm{PS}=10$ mW and $P_\mathrm{PA}=100$ mW \cite{rappaport2014millimeter}.
\begin{figure}[htbp]
\centering\includegraphics[width=8cm]{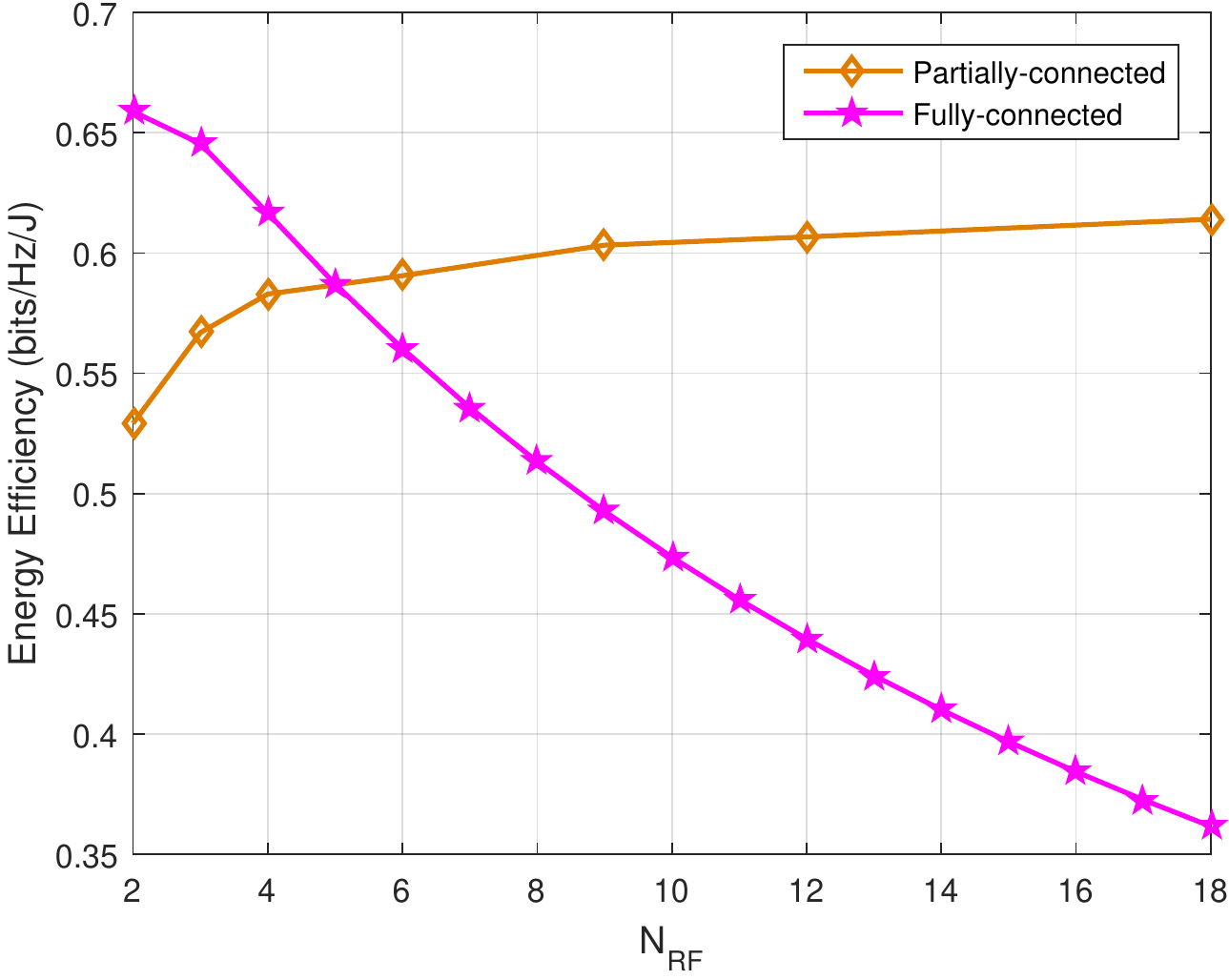}
\caption{Energy efficiency of the fully-connected and partially-connected structures when $N_s=2$, $N_\mathrm{RF}^t=N_\mathrm{RF}^r=N_\mathrm{RF}$ and $\mathrm{SNR}=0$ dB.}\label{f3}
\end{figure}
The simulation results are shown in Fig. \ref{f3}, which shows substantially different behaviors for the two structures. Since the number of phase shifters scales linearly with $N_\mathrm{RF}^t$ and $N_t$ in the fully-connected structure, the power consumption will increase substantially when increasing $N_\mathrm{RF}^t$. As shown in Fig. \ref{f2}, however, the spectral efficiency achieved by the proposed MO-AltMin algorithm is sufficiently close or exactly equal to the optimal digital one, and will not increase further as $N_\mathrm{RF}$ increases. Based on these two facts, the power consumption grows much faster than the spectral efficiency, which gives rise to the dramatic decrease of the energy efficiency.

For the partially-connected structure, as the number of phase shifters is independent of $N_\mathrm{RF}^t$, the dominant part of total power consumption remains almost unchanged over the investigated range of RF chain numbers. Meanwhile, the spectral efficiency will gradually approach the optimal digital precoder when increasing $N_\mathrm{RF}^t$.
The improvement of the spectral efficiency and the almost unchanged power consumption together account for the rise in the energy efficiency when $N_\mathrm{RF}^t$ goes up in the partially-connected structure.

More importantly, Fig. \ref{f3} shows that there is an intersection point, i.e., when $N_\mathrm{RF}=5$, of the energy efficiency for two hybrid precoding structures. In particular, the fully-connected structure enjoys higher energy efficiency with a small number of RF chains, while the partially-connected one is more energy efficient when a relatively large number of RF chains are implemented at transceivers. This phenomenon will offer valuable insights for the RF chain implementation in hybrid precoding. As shown in Fig. \ref{f2}, the fully-connected structure can approach the performance of the optimal digital precoder when the number of RF chains is slightly larger than that of the data streams. Therefore, there is no need to implement more RF chains considering the energy efficiency. On the other hand, with a low-complexity hardware implementation, it is beneficial for the partially-connected structure to leverage the larger size of RF chains to improve both spectral and energy efficiency.

\subsection{Low-Complexity Design for the Fully-connected Structure}
As mentioned in Section \ref{issue}, there are two intermediate steps taken for developing the PE-AltMin algorithm, i.e., constructing the digital precoder with orthogonal columns and replacing the objective function with an upper bound. Firstly, we test the influence of the additional constraint on the digital precoder, for which we eliminate the impact of the second step by setting the numbers of RF chains and data streams to be equal, i.e., to make the bound tight.
\begin{figure}[htbp]
\centering\includegraphics[width=8cm]{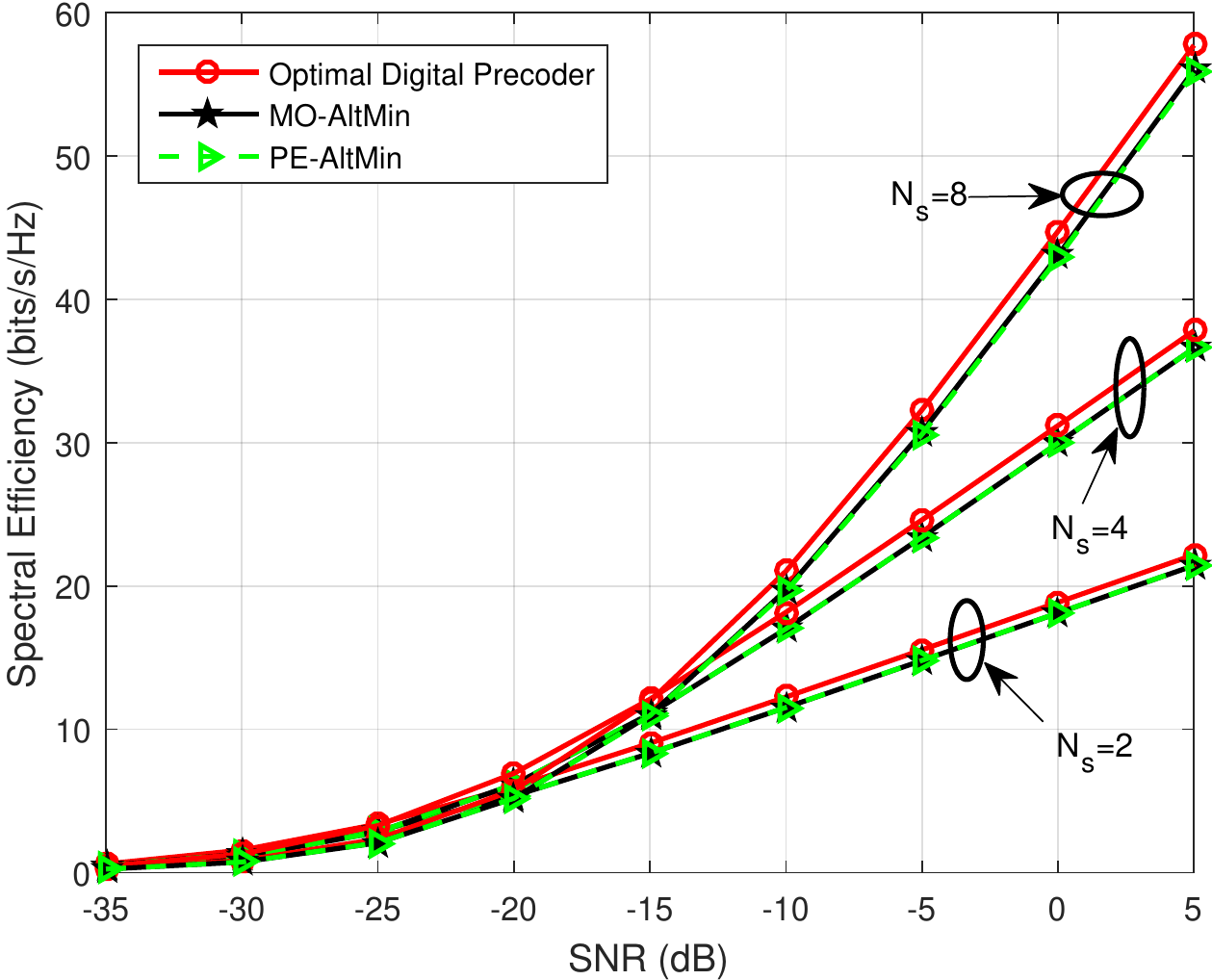}
\caption{Spectral efficiency achieved by the MO-AltMin and PE-AltMin algorithms given $N_\mathrm{RF}^t=N_\mathrm{RF}^r=N_\mathrm{RF}=N_s$.}\label{n2}
\end{figure}

Fig. \ref{n2} plots the spectral efficiency achieved by both the MO-AltMin and PE-AltMin algorithms when there are 2, 4 and 8 data streams transmitted.
We see that the curves of the low-complexity PE-AltMin algorithm nearly coincide with those of the MO-AltMin algorithm. This phenomenon implies that the orthogonal column structure of the digital precoder has negligible impact on spectral efficiency, which justifies the rationality of the digital precoder design in Section \ref{issue}. Fig. \ref{n2} also indicates that we can achieve the performance of the high complexity MO-AltMin algorithm by adopting the low-complexity PE-AltMin algorithm when $N_\mathrm{RF}=N_s$. Under this system setting, the PE-AltMin algorithm serves as an excellent candidate for the hybrid precoder design, achieving both good performance and low complexity. On the contrary, the OMP algorithm works poorly when $N_\mathrm{RF}=N_s$.
\begin{figure}[t]
\centering\includegraphics[width=8cm]{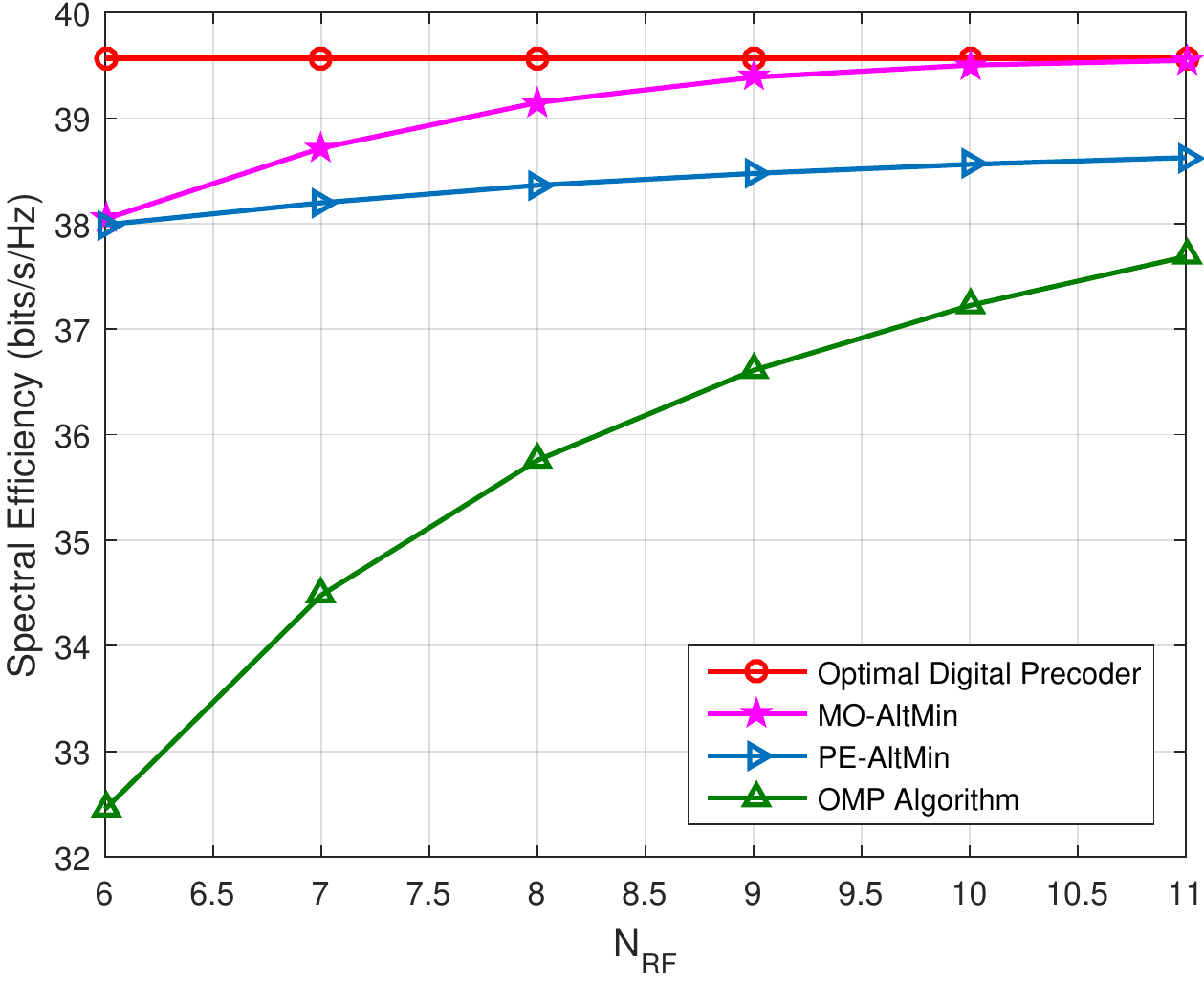}
\caption{Spectral efficiency achieved by different precoding algorithms given $N_s=6$, $N_\mathrm{RF}^t=N_\mathrm{RF}^r=N_\mathrm{RF}$ and $\mathrm{SNR}=0$ dB.}\label{n3}
\end{figure}

Next we investigate the impact of the number of RF chains. Fig. \ref{n3} compares different algorithms assuming 6 data streams are transmitted. Since the optimal solution in the $N_\mathrm{RF}^t\ge2N_s$ region has been fully developed in \cite{zhang2014achieving}, here we focus on the remaining region, i.e., $N_\mathrm{RF}^t=N_\mathrm{RF}^r=N_\mathrm{RF}\in[6,11]$. From Fig. \ref{n3}, we find that the PE-AltMin algorithm has a small gap compared with the MO-AltMin algorithm. That is because the low-complexity algorithm tries to minimize an upper bound instead of the original objective function. As explained in Section \ref{issue}, the upper bound is tight when $N_\mathrm{RF}=N_s$ and gets looser when $N_\mathrm{RF}$ increases, which determines the gap between the MO-AltMin and PE-AltMin algorithms. However, the spectral efficiency provided by the PE-AltMin algorithm is far higher than the existing OMP algorithm, especially when the RF chain number is small. 
With the proposed AltMin algorithms, we see that the performance of the fully-connected structure approaches that of the fully digital precoder when the number of RF chains is comparable with the number of data streams, which cannot be revealed from the OMP algorithm.

\subsection{Hybrid Precoding in MmWave MIMO-OFDM Systems}
In this part, we will show the performance of the proposed AltMin algorithms when applied to mmWave MIMO-OFDM systems. We assume that the number of subcarriers is $K=128$.
\begin{figure}[tbp]
\centering\includegraphics[width=8cm]{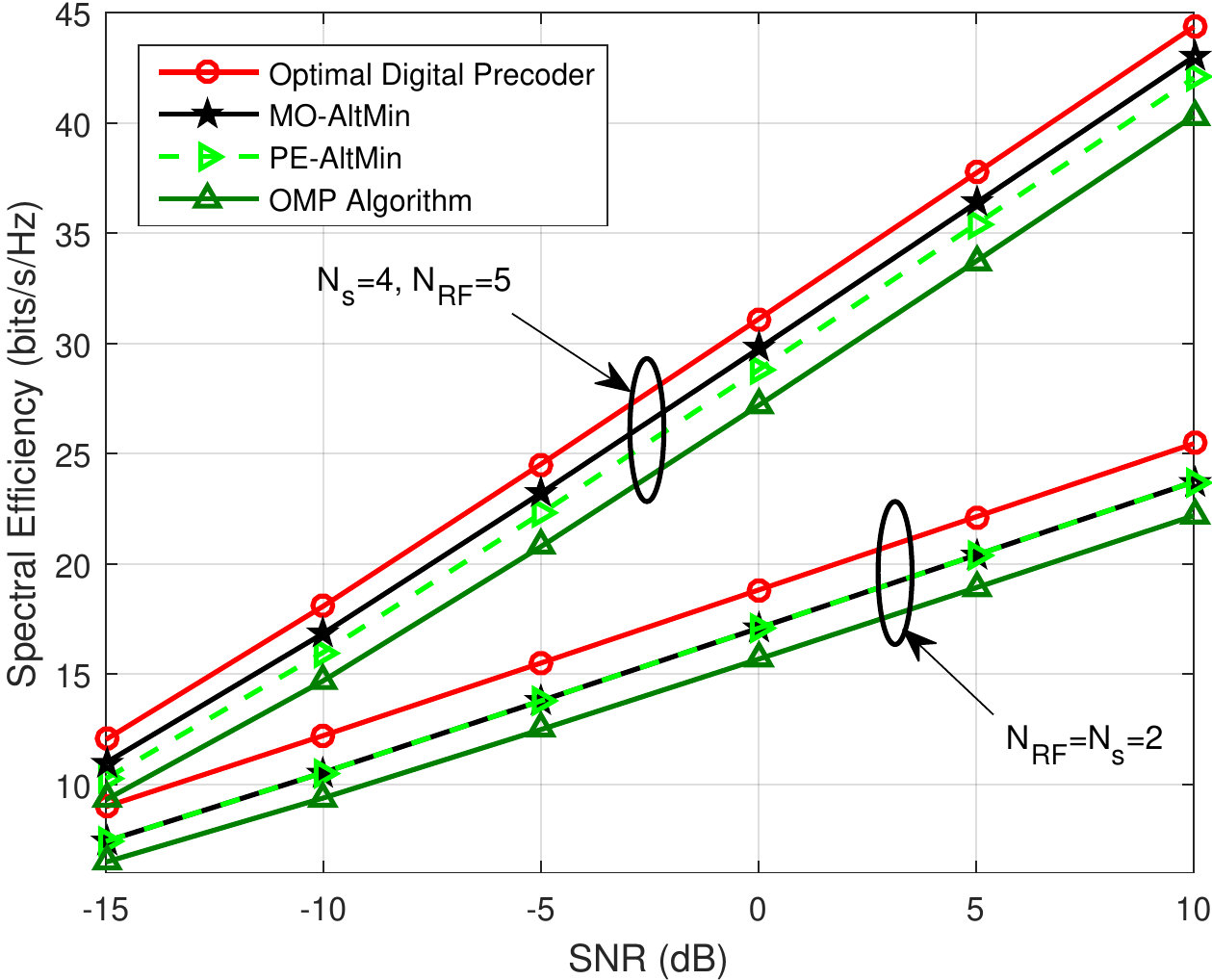}
\caption{Spectral efficiency achieved by different precoding algorithms in mmWave MIMO-OFDM systems given $N_\mathrm{RF}^t=N_\mathrm{RF}^r=N_\mathrm{RF}$.}\label{simulation6}
\end{figure}

Fig. \ref{simulation6} plots the spectral efficiency achieved by the MO-AltMin and PE-AltMin algorithms compared to the OMP-based method proposed in \cite{6884253}. It turns out that the MO-AltMin algorithm always has the highest spectral efficiency under different system parameters. With a large number of RF chains, similar to what we have observed in Fig. \ref{n3}, the MO-AltMin algorithm will quickly get close to the performance of the optimal digital precoder. Interestingly, we discover that the low-complexity PE-AltMin algorithm can achieve almost the same spectral efficiency as that of the MO-AltMin algorithm when the numbers of RF chains and data streams are equal. This phenomenon is the same as that in narrowband systems, and it demonstrates that the extra orthogonality constraint on the digital precoder also has negligible impact on spectral efficiency in mmWave OFDM systems. It is also found in Fig. \ref{simulation6} that the PE-AltMin algorithm outperforms the OMP-based method in mmWave OFDM systems. This indicates that the PE-AltMin algorithm can serve as an outstanding candidate for the low-complexity hybrid precoding, both in narrowband and broadband OFDM systems, when the transceivers only have limited RF chains available.

\section{Conclusions}
Built on the principle of alternating minimization, in this paper, we proposed an innovative design methodology for hybrid precoding in mmWave MIMO systems. 
Effective algorithms were proposed for the fully-connected and partially-connected hybrid precoding structures, and simulation results helped reveal the following valuable design insights:
\begin{itemize}
\item The hybrid precoders with the fully-connected structure can approach the performance of the fully digital precoder when the number of RF chains is slightly larger than the number of data streams. Considering the increasing cost and power consumption, there is no need to further increase the number of RF chains.
\item For the partially-connected structure, in terms of spectral efficiency, hybrid precoders provide substantial gains over analog beamforming. Furthermore, it is profitable to implement a relatively large number of RF chains, in order to enhance both spectral and energy efficiency.
\end{itemize}
Finally, our results have clearly demonstrated the effectiveness of alternating minimization in designing hybrid precoders in mmWave MIMO systems. It will be interesting to extend the alternating minimization techniques to other hybrid precoder design problems, as well as to consider the hybrid precoder design combined with channel training and feedback. Also a finer convergence analysis and optimality characterization of the proposed algorithms will require further investigation.


%

\appendices
\section{Formulation of the Homogeneous QCQP Problem}\label{appA}
The original problem is a non-homogeneous QCQP problem
\begin{equation}\label{38}
\begin{aligned}
&\underset{\mathbf{F}_\mathrm{BB}}{\mathrm{minimize}} && \left\Vert \mathbf{F}_\mathrm{opt}-\mathbf{F}_\mathrm{RF}\mathbf{F}_\mathrm{BB}\right\Vert _F^2\\
&\mathrm{subject\thinspace to} &&{\left\| \bf{F_{\rm{BB}}} \right\|}_F^2=\frac{N_\mathrm{RF}^tN_s}{N_t}.
\end{aligned}
\end{equation}
According to the vectorization property, the objective function in \eqref{38} can be rewritten as
\begin{equation}
\begin{split}
\left\Vert \mathbf{F}_\mathrm{opt}-\mathbf{F}_\mathrm{RF}\mathbf{F}_\mathrm{BB}\right\Vert _F^2&=
\left\Vert \rm{vec}(\mathbf{F}_{\mathrm{opt}} - \mathbf{F}_{\mathrm{RF}}\mathbf{F}_{\mathrm{BB}})\right\Vert _2^2\\
&=\left\Vert\rm{vec}(\mathbf{F}_{\mathrm{opt}}) -\rm{vec}( \mathbf{F}_{\mathrm{RF}}\mathbf{F}_{\mathrm{BB}})\right\Vert_2^2\\
&=\left\Vert\mathrm{vec}(\mathbf{F}_{\mathrm{opt}}) - (\mathbf{I}_{N_s}\otimes\mathbf{F}_{\mathrm{RF}})\rm{vec}(\mathbf{F}_{\mathrm{BB}})\right\Vert_2^2.
\end{split}
\end{equation}
To simplify the notation, denote $\mathbf{f}=\mathrm{vec}\left(\Fopt\right)$, $\mathbf{b}=\mathrm{vec}\left(\FBB\right)$ and $\mathbf{E}=\mathbf{I}_{N_s}\otimes\FRF$.
In order to apply SDR, we introduce an auxiliary variable $t$ to homogenize the original problem as
\begin{equation}\label{40}
\begin{aligned}
&\underset{\mathbf{b}}{\mathrm{minimize}} && \|t\mathbf{f} - \mathbf{Eb}\|_2^2\\
&\mathrm{subject\thinspace to}&&
\begin{cases}
\|\mathbf{b}\|_2^2=\frac{N_\mathrm{RF}^tN_s}{N_t}\\
t^2=1.
\end{cases}
\end{aligned}
\end{equation}

The objective function in \eqref{40} can be further rewritten as
\begin{equation}
\begin{split}
\|t\mathbf{f} - \mathbf{Eb}\|_2^2=\left[ {\begin{array}{*{20}{c}}
\mathbf{b}^H&{t}
\end{array}} \right]\left[ {\begin{array}{*{20}{c}}
\mathbf{E}^H\mathbf{E}&
-\mathbf{E}^H\mathbf{f}\\
-\mathbf{f}^H\mathbf{E}&
\mathbf{f}^H\mathbf{f}
\end{array}} \right]\left[ {\begin{array}{*{20}{c}}
\mathbf{b}\\{t}
\end{array}} \right].\\
\end{split}
\end{equation}
Furthermore, the two constraints in \eqref{40} can also be manipulated as
\begin{equation}
\begin{split}
\|\mathbf{b}\|_2^2=\left[ {\begin{array}{*{20}{c}}
\mathbf{b}^H&{t}
\end{array}} \right]\left[ {\begin{array}{*{20}{c}}
\mathbf{I}_{N_\mathrm{RF}^tN_s}&
\mathbf{0}\\
\mathbf{0}&
0
\end{array}} \right]\left[ {\begin{array}{*{20}{c}}
\mathbf{b}\\{t}
\end{array}} \right]=\frac{N_\mathrm{RF}^tN_s}{N_t},\\
\end{split}
\end{equation}

\begin{equation}
t^2=
\left[ {\begin{array}{*{20}{c}}
{\mathbf{b}^H}&{t}
\end{array}} \right]\left[ {\begin{array}{*{20}{c}}
{\mathbf{0}_{N_\mathrm{RF}^tN_s}}&{\mathbf{0}}\\
{\mathbf{0}}&{1}
\end{array}} \right]\left[ {\begin{array}{*{20}{c}}
{\mathbf{b}}\\
{t}
\end{array}} \right]=1.
\end{equation}

Consequently, if we define
\begin{equation*}
\begin{split}
&\mathbf{y}=\left[ {\begin{array}{*{20}{c}}
{\mathbf{b}}\\
{t}
\end{array}} \right],\mathbf{Y}=\mathbf{yy}^H,\mathbf{C}=\left[ {\begin{array}{*{20}{c}}
\mathbf{E}^H\mathbf{E}&
-\mathbf{E}^H\mathbf{f}\\
-\mathbf{f}^H\mathbf{E}&
\mathbf{f}^H\mathbf{f}
\end{array}} \right],
\\
&\mathbf{A}_1=\left[ {\begin{array}{*{20}{c}}
{\mathbf{I}_{N_\mathrm{RF}^tN_s}}&{\mathbf{0}}\\
{\mathbf{0}}&{0}
\end{array}} \right],
\mathbf{A}_2=\left[ {\begin{array}{*{20}{c}}
{\mathbf{0}_{N_\mathrm{RF}^tN_s}}&{\mathbf{0}}\\
{\mathbf{0}}&{1}
\end{array}} \right],
\end{split}
\end{equation*}
the original problem \eqref{38} can be reformulated as
\begin{equation}
\begin{aligned}
&\underset{\mathbf{Y}\in\mathbb{H}^n}{\mathrm{minimize}} && \mathrm{Tr}(\mathbf{CY})\\
&\mathrm{subject\thinspace to}&&
\begin{cases}
\mathrm{Tr}(\mathbf{A}_1\mathbf{Y})=\frac{N_\mathrm{RF}^tN_s}{N_t}\\
\mathrm{Tr}(\mathbf{A}_2\mathbf{Y})=1\\
\mathbf{Y}\succeq 0,\;\mathrm{rank}(\mathbf{Y})=1.
\end{cases}
\end{aligned}
\end{equation}
%
%

\ifCLASSOPTIONcaptionsoff
  \newpage
\fi



%
\bibliographystyle{IEEEtran}
\bibliography{bare_jrnl}
%

\begin{IEEEbiography}
[{\includegraphics[width=1in,height=1.25in,clip,keepaspectratio]{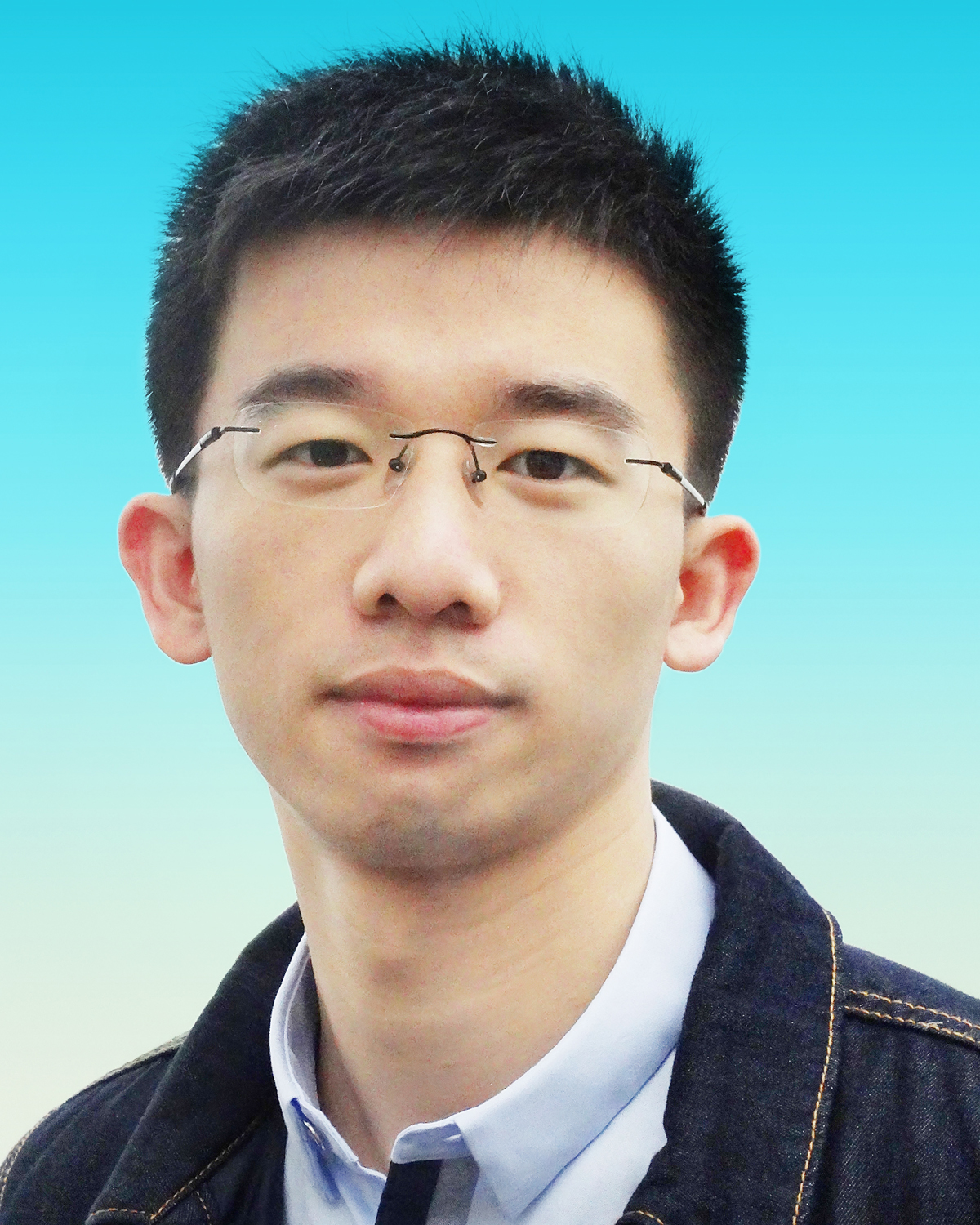}}]{Xianghao Yu}
(S'15) received the B.Eng. degree in Information Engineering from Southeast University (SEU), Nanjing, China, in 2014. He is currently working towards the Ph.D. degree in Electronic and Computer Engineering at the Hong Kong University of Science and Technology (HKUST), under the supervision of Prof. Khaled B. Letaief. His research interests include millimeter wave communications, MIMO systems, mathematical optimization, and stochastic geometry.
\end{IEEEbiography}
\begin{IEEEbiography}
[{\includegraphics[width=1in,height=1.25in,clip,keepaspectratio]{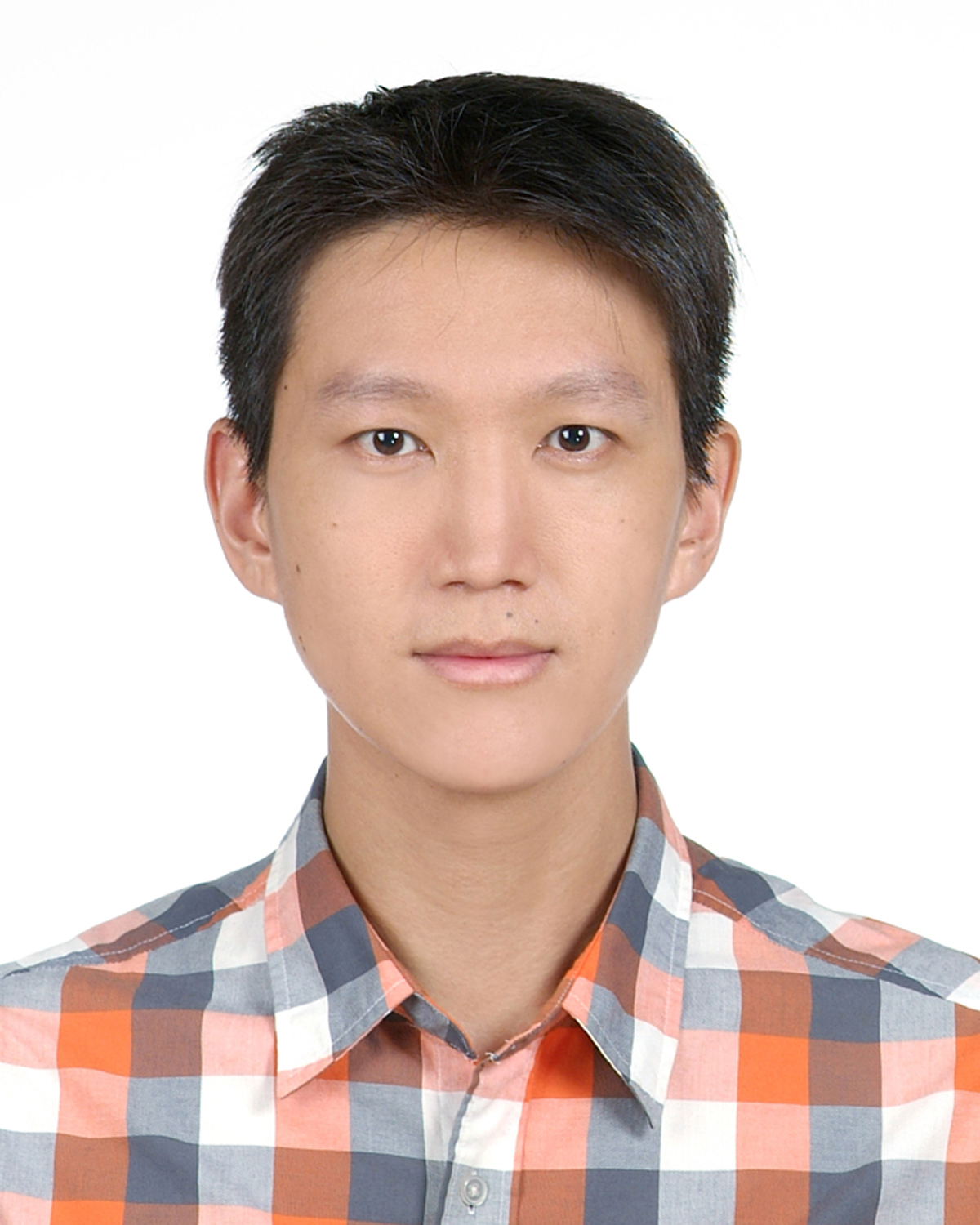}}]{Juei-Chin Shen}
(S'10--M'14) received the Ph.D. degree in communication engineering from University of Manchester, United Kingdom, in 2013. He was with the Hong Kong University of Science and Technology (HKUST) from 2013 to 2015 as a Research Associate. Since 2015, he has been with Mediatek Inc. as a Senior Engineer, working on the research and standardization of future wireless technologies. He is a co-recipient of IEEE PIMRC 2014 Best Paper Award. His research interests include compressed sensing, massive MIMO systems, and millimeter-wave communications.
\end{IEEEbiography}

\begin{IEEEbiography}
[{\includegraphics[width=1in,height=1.25in,clip,keepaspectratio]{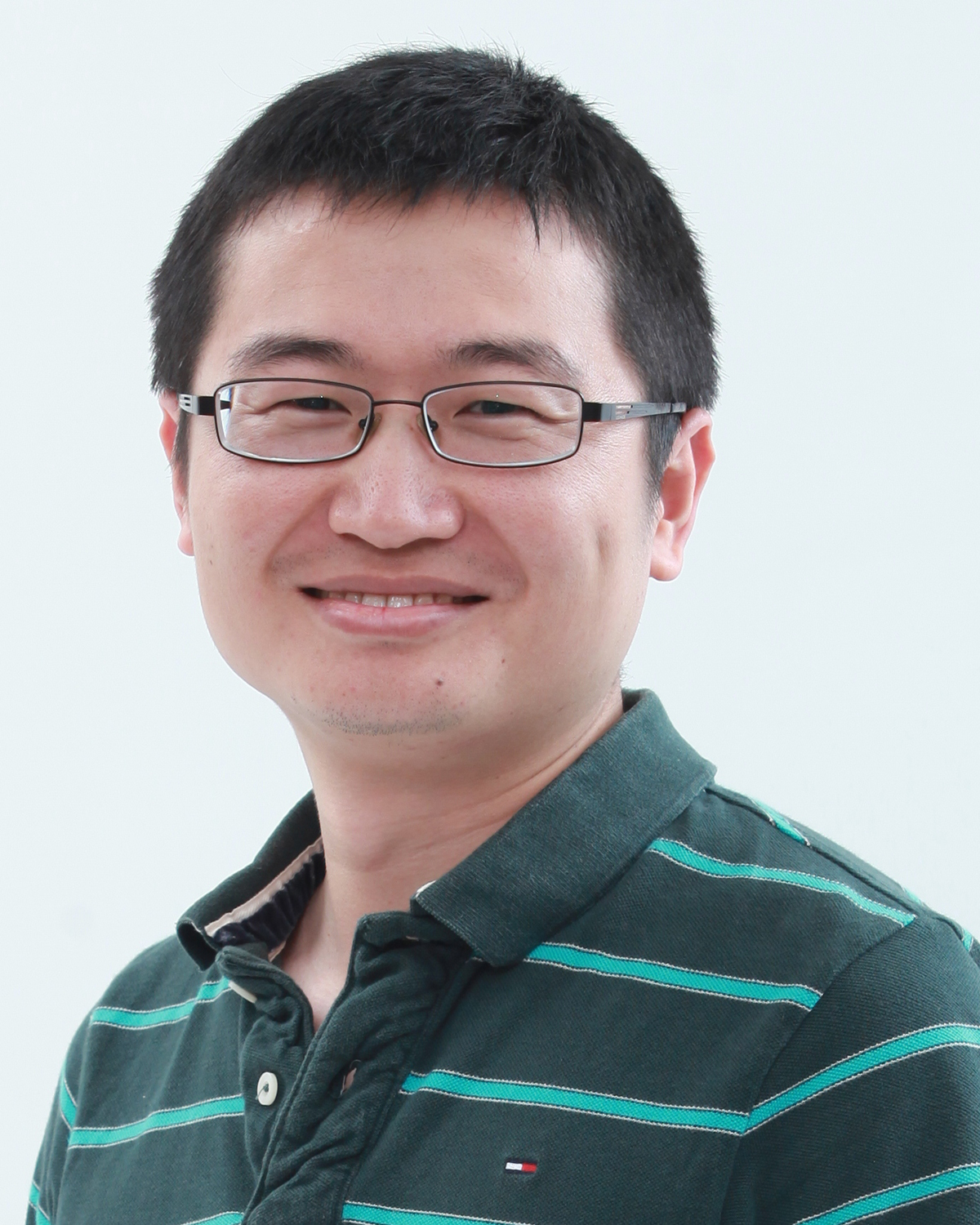}}]{Jun Zhang}
(S'06--M'10--SM'15) received the B.Eng. degree in Electronic Engineering from the University of Science and Technology of China in 2004, the M.Phil. degree in Information Engineering from the Chinese University of Hong Kong in 2006, and the Ph.D. degree in Electrical and Computer Engineering from the University of Texas at Austin in 2009. He is currently a Research Assistant Professor in the Department of Electronic and Computer Engineering at the Hong Kong University of Science and Technology (HKUST). Dr. Zhang co-authored the book \emph{Fundamentals of LTE} (Prentice-Hall, 2010). He received the 2014 Best Paper Award for the EURASIP Journal on Advances in Signal Processing, and the PIMRC 2014 Best Paper Award. He is an Editor of IEEE Transactions on Wireless Communications, and served as a MAC track co-chair for IEEE WCNC 2011. His research interests include wireless communications and networking, green communications, and signal processing.
\end{IEEEbiography}

\begin{IEEEbiography}
[{\includegraphics[width=1in,height=1.25in,clip,keepaspectratio]{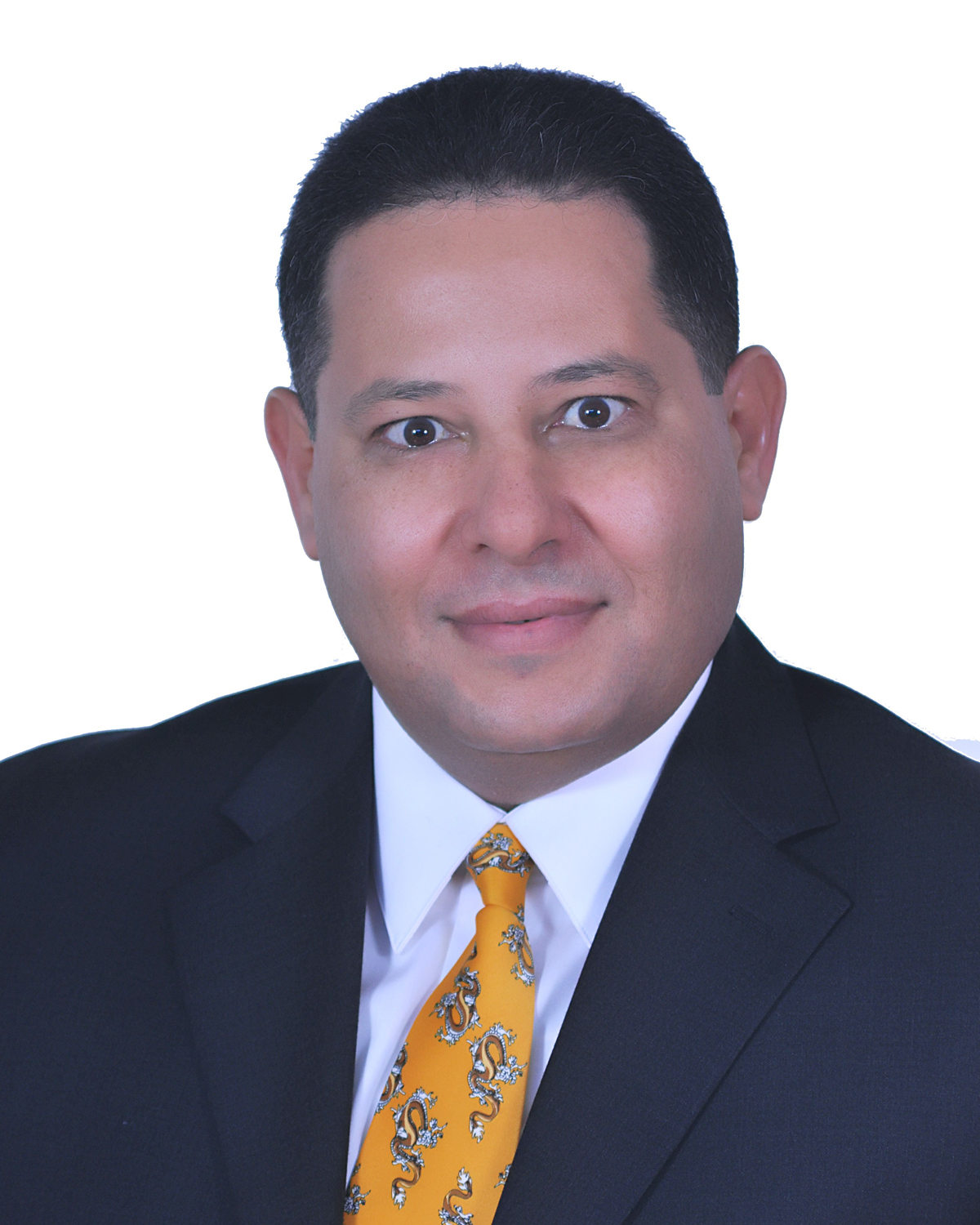}}]{Khaled B. Letaief}
(S'85--M'86--SM'97--F'03) received the B.S. degree \emph{with distinction} in Electrical Engineering (1984) from Purdue University, USA. He has also received the M.S. and Ph.D. Degrees in Electrical Engineering from Purdue University in 1986 and 1990, respectively.

From 1990 to 1993, he was a faculty member at the University of Melbourne, Australia. He has been with the Hong Kong University of Science \& Technology. While at HKUST, he has held numerous administrative positions, including the Head of the Department of Electronic and Computer Engineering, Director of the Center for Wireless IC Design, Director of Huawei Innovation Laboratory, and Director of the Hong Kong Telecom Institute of Information Technology. He has also served as Chair Professor and Dean of HKUST School of Engineering. Under his leadership, the School of Engineering has dazzled in international rankings (\emph{ranked \#14 in the world in 2015} according to QS World University Rankings).

From September 2015, he joined Hamad Bin Khalifa University (HBKU) as Provost to help establish a research-intensive university in Qatar in partnership with strategic partners that include Northwestern University, Carnegie Mellon University, Cornell, and Texas A\&M.

Dr. Letaief is a world-renowned leader in wireless communications and networks. In these areas, he has over 500 journal and conference papers and given invited keynote talks as well as courses all over the world. He has made 6 major contributions to IEEE Standards along with 13 patents including 11 US patents.

He served as consultants for different organizations and is the founding Editor-in-Chief of the prestigious \emph{IEEE Transactions on Wireless Communications}. He has served on the editorial board of other prestigious journals including the \emph{IEEE Journal on Selected Areas in Communications -- Wireless Series} (as Editor-in-Chief). He has been involved in organizing a number of major international conferences.

Professor Letaief has been a dedicated educator committed to excellence in teaching and scholarship. He received the \emph{Mangoon Teaching Award} from Purdue University in 1990; HKUST Engineering Teaching Excellence Award (4 times); and the Michael Gale Medal for Distinguished Teaching (\emph{Highest university-wide teaching award} at HKUST).

He is also the recipient of many other distinguished awards including 2007 \emph{IEEE Joseph LoCicero} Publications Exemplary Award; 2009 IEEE Marconi Prize Award in Wireless Communications; 2010 Purdue University Outstanding Electrical and Computer Engineer Award; 2011 IEEE Harold Sobol Award; 2011 IEEE Wireless Communications Technical Committee Recognition Award; and 11 IEEE Best Paper Awards.

He had the privilege to serve IEEE in many leadership positions including IEEE ComSoc Vice-President, IEEE ComSoc Director of Journals, and member of IEEE Publications Services and Products Board, IEEE ComSoc Board of Governors, IEEE TAB Periodicals Committee, and IEEE Fellow Committee.

Dr. Letaief is a Fellow of IEEE and a Fellow of HKIE. He is also recognized by Thomson Reuters as an \emph{ISI Highly Cited Researcher}.
\end{IEEEbiography}
%
%




\end{document}